\documentclass{IEEEtran} 

\usepackage{graphicx,color}
\usepackage{dcolumn}
\usepackage{bm}
\usepackage{mathbbol}
\usepackage{amsmath,amssymb}  
\usepackage[latin9]{inputenc} 
\usepackage{rotating}
\usepackage[table]{xcolor}

\makeatletter

\newtheorem{prop}{Proposition}
\newtheorem{defin}{Definition}
\newtheorem{thm}{Theorem}
\newtheorem{cor}{Corollary}

\newcommand{\ket}[1]{|#1\rangle}
\newcommand{\bra}[1]{\langle #1|}

\newcommand{\Hi}{\mathcal{H}}

\newcommand{\supp}{\textrm{supp}}

\newcommand{\beq}{\begin{equation}}
\newcommand{\eeq}{\end{equation}}
\newcommand{\bea}{\begin{eqnarray}}
\newcommand{\eea}{\end{eqnarray}}
\newcommand{\beqa}{\begin{eqnarray}}
\newcommand{\eeqa}{\end{eqnarray}}
\newcommand{\beqan}{\begin{eqnarray*}}
\newcommand{\eeqan}{\end{eqnarray*}}

\def\smallfrac#1#2{{\textstyle\frac{#1}{#2}}}

\newcommand{\R}{\mathbb{R}}

\newcommand{\Li}{\mathcal{L}}

\newcommand{\diag}{\textrm{diag}}
\newcommand{\spec}{\textrm{sp}}
\newcommand{\tr}{\textrm{trace}}
\newcommand{\real}{\mathfrak{Re}}

\renewcommand{\Re}{\mathfrak{Re}}

\newcommand{\bounded}[1]{\mathcal{B}(\mathcal{H}_{#1})}

\newcommand{\dopset}[1]{\mathfrak{D}(\mathcal{H}_{#1})}
\newcommand{\initset}[2]{\mathfrak{I}_{#1}(\mathcal{H}_{#2})}
\newcommand{\herm}[1]{\mathfrak{h}(\mathcal{H}_{#1})}

\newcommand{\hatLi}{\mathcal{\hat L}}
\def \identity{\mathbf{1}}
\def \vec{\textrm{vec}}

\def \sign{\mathrm{sign}}
\def \span{\mathrm{span}}

\begin{document}

\title{Hamiltonian Control of Quantum Dynamical Semigroups: 
Stabilization and Convergence Speed}

\author{Francesco Ticozzi, Riccardo Lucchese, Paola Cappellaro and
Lorenza Viola

\thanks{Francesco Ticozzi is with the Dipartimento di Ingegneria
dell'Informazione, Universit\`a di Padova, via Gradenigo 6/B, 35131
Padova, Italy (email: \mbox{ticozzi@dei.unipd.it}).}

\thanks{Riccardo Lucchese is with the Dipartimento di Ingegneria
dell'Informazione, Universit\`a di Padova, via Gradenigo 6/B, 35131
Padova, Italy (email: \mbox{lucchese@dei.unipd.it}).}

\thanks{Paola Cappellaro is with Department of Nuclear Science and
Engineering, Massachusetts Institute of Technology, 77 Massachusetts
Avenue, Cambridge, MA 02139, USA (email: \mbox{pcappell@mit.edu}).}

\thanks{Lorenza Viola is with Department of Physics and Astronomy,
Dartmouth College, 6127 Wilder Laboratory, Hanover, NH 03755, USA
(email: \mbox{lorenza.viola@dartmouth.edu}).}

\thanks{F. T. acknowledges hospitality from the Physics and Astronomy
Department at Dartmouth College, where this work was performed, and
support from the University of Padova under the QUINTET project of the
Department of Information Engineering, and the QFUTURE and
CPDA080209/08 grants. }
}


\maketitle

\begin{abstract}
We consider finite-dimensional Markovian open quantum systems, and
characterize the extent to which time-independent Hamiltonian control
may allow to stabilize a target quantum state or subspace and optimize
the resulting convergence speed.  For a generic Lindblad master
equation, we introduce a dissipation-induced decomposition of the
associated Hilbert space, and show how it serves both as a tool to
analyze global stability properties for given control resources and as
the starting point to synthesize controls that ensure rapid
convergence.  The resulting design principles are illustrated in
realistic Markovian control settings motivated by quantum information
processing, including quantum-optical systems and nitrogen-vacancy
centers in diamond.
\end{abstract}

\section{Introduction}

Devising effective strategies for stabilizing a desired quantum state
or subsystem under general dissipative dynamics is an important
problem from both a control-theoretic and quantum engineering
standpoint.  Significant effort has been recently devoted, in
particular, to the paradigmatic class of Markovian open quantum
systems, whose (continuous-time) evolution is described by a quantum
dynamical semigroup \cite{alicki-lendi}.  Building on earlier
controllability studies \cite{altafini-markovian,altafini-open},
Markovian stabilization problems have been addressed in settings
ranging from the preparation of complex quantum states in multipartite
systems to the synthesis of noiseless quantum information encodings by
means of open-loop Hamiltonian control and reservoir engineering as
well as quantum feedback
\cite{ticozzi-QDS,ticozzi-markovian,schirmer-markovian,Kraus-entanglement,ticozzi-generic,schirmer-3}.
While a number of rigorous results and control protocols have emerged,
the continuous progress witnessed by laboratory quantum technologies
makes it imperative to develop theoretical approaches which attempt to
address practical constraints and limitations to the extent possible.
 
In this work, we focus on the open-loop stability properties of
quantum semigroup dynamics that is solely controlled in terms of {\em
time-independent Hamiltonians}, with a twofold motivation in mind: (i)
determining under which conditions a desired target state or subspace
may be stabilizable given limited control resources; (ii)
characterizing how Hamiltonian control influences the asymptotic speed
of convergence to the target set.  A number of analysis tools are
obtained to this end.  On the one hand, we introduce the notion of a
{\em dissipation-induced decomposition} of the system Hilbert space as
a canonical way to represent and study Markovian dissipative dynamics
(Sec. \ref{sectionDID}). A constructive algorithm for determining such
a (unique) decomposition is presented, as well as an enhanced
algorithm that finds which control inputs, if any, can ensure
convergence under fairly general constraints (Sec. \ref{constraints}).
On the other hand, we present two approaches to analyze the
convergence of the semigroup to the target set: the first, which is
system-theoretic in nature, offers in principle a quantitative way to
computing the asymptotic speed of convergence (or Lyapunov exponent)
from a certain reduced dynamical matrix (Sec. \ref{speednum}; the
second, which builds directly on the above dissipation-induced
decomposition and we term {\em connected basins approach}, offers a
qualitative way to estimating the convergence speed and designing
control in situations where exact analytical or numerical methods may
be impractical (Sec. \ref{basins}).  By using these tools, we show how
a number of fundamental issues related to role of the Hamiltonian in
the convergence of quantum dynamical semigroups can be tackled, also
leading to further physical insight on the interplay between coherent
control and dissipation \cite{bartana-tannor}.  A number of physically
motivated examples are discussed at length in Sec. \ref{application},
demonstrating how our approach can be useful in realistic quantum
control scenarios.

\section{Quantum dynamical semigroups}

\subsection{Open-loop controlled QDS dynamics}

Throughout this work, we shall consider a finite-dimensional open
quantum system with associated complex Hilbert space $\Hi,$ with
$\dim(\Hi)=n$. Using Dirac's notation \cite{sakurai}, we denote
vectors in $\Hi$ by $\ket{\psi},$ and linear functionals in the dual
$\Hi^\dag \simeq \Hi$ by $\bra{\phi}$. Let in addition $\bounded{}$ be
the set of linear operators on $\Hi,$ with $\herm{}$ being the
Hermitian ones, and $\dopset{}\subset \herm{}$ the trace-one, positive
semidefinite operators (or {\em density operators}), which physically
represent the {\em states} of the quantum system.  The class of
dynamics we consider are one-parameter quantum dynamical semigroups
(QDSs), that is, continuous families of trace-preserving completely
positive maps $\{{\cal T}_t\}_{t\geq 0}$ that enjoy a forward (Markov)
composition law.  The evolution of states $\rho \in \dopset{}$ is then
governed by a master equation
\cite{lindblad,qds,gorini-k-s,alicki-lendi} of the Lindblad form:
\begin{equation} 
\label{eq:lindblad}
\dot\rho=\Li[\rho] = -\frac{i}{\hbar} [H,\rho] + \sum_{k=1}^p \Big(L_k
\rho L_k^\dag -\frac{1}{2} \{L_k^\dag L_k, \rho\}\Big),
\end{equation}
where the effective Hamiltonian $H\in \herm{}$ and the noise operators
$\{L_k\} \subset \bounded{}$ describe, respectively, the coherent
(unitary) and dissipative (non-unitary) contributions to the dynamics.
In what follows, $\hbar=1$ unless otherwise stated.

We focus here on control scenarios where the Hamiltonian $H$ of
the system can be tuned through suitable control inputs, that is,  
\begin{equation} 
\label{eq:ham}
H(u)=H_0+\sum_{j=1}^\nu H_j u_j,
\end{equation}
where $H_0=H_0^\dag$ represents the free (internal) system
Hamiltonian, and the controls $u_j\in\R$ modify the dynamics through
the Hamiltonians $H_j=H_j^\dag$. In particular, we are interested in
the case of constant controls $u_j$ taking values in some (possibly
open) interval ${\cal C}_j\subseteq \bar\R$.  The set of admissible
control choices is then a subset ${\cal C}\subseteq \R^\nu.$

\subsection{Stable subspaces for QDS dynamics} 

We begin by recalling some relevant definitions and results of the
linear-algebraic approach to stabilization of QDS developed in
\cite{ticozzi-QDS,ticozzi-markovian,ticozzi-generic,bolognani-arxiv,schirmer-markovian}. Consider
an orthogonal decomposition of the Hilbert space $\Hi := \Hi_S \oplus
\Hi_R,$ with $\dim(\Hi_S)=m \leq n.$ Let $\{\ket{s_i}\}$ and
$\{\ket{r_j}\}$ be orthonormal sets spanning $\Hi_S$ and $\Hi_R$
respectively.  The (ordered) basis
$\{\ket{s_1},\ldots,\ket{s_m},\ket{r_1},\ldots,\ket{r_{n-m}}\}$
induces the following block structure on the matrix representation of
an arbitrary operator $X \in \bounded{}$:
\begin{equation} 
\label{eq-block-structure}
X = \begin{bmatrix}X_S & X_P \\ X_Q & X_R 
\end{bmatrix}.
\end{equation}
Let in addition $\supp(X):= \ker(X)^\bot$. It will be useful to
introduce a compact notation for sets of states with support contained
in a given subspace:
$$\initset{S}{}:=\Big\{\rho\in\dopset{}\,|\,\rho=\begin{bmatrix}\rho_S
                    & 0 \\ 0 & 0\end{bmatrix}, \:\rho_S\in\dopset{S}
                    \Big\}.$$

As usual in the study of dynamical systems, we say that a set of
states ${\cal W}$ is {\em invariant} for the dynamics generated by
$\Li$ if arbitrary trajectories originating in ${\cal W}$ at $t=0$
remain confined to ${\cal W}$ at all positive times.  Henceforth, with
a slight abuse of terminology, we will say that a {\em subspace $\Hi_S
\subset \Hi$ is $\Li$-invariant} (or simply invariant) when
$\initset{S}{}$ is invariant for the dynamics generated by $\Li$.

An algebraic characterization of ``subspace-invariant'' QDS generators
is provided by the following Proposition (the proof is given in
\cite{ticozzi-QDS}, in the more general {\em subsystem} case):

\begin{prop}[{$S$-Subspace Invariance}] 
\label{pro-markovian-invariant-generators} 
Consider a QDS on $\Hi=\Hi_S \oplus \Hi_R$, and let the generator
$\Li$ in \eqref{eq:lindblad} be associated to an Hamiltonian $H$ and a
set of noise operators $\{L_k\}$. Then $\Hi_S$ is invariant if and
only if the following conditions hold:
\begin{equation} 
\label{invarianceconditions}
  \left \{ \begin{aligned} &iH_P-\frac{1}{2}\sum_k L_{S,k}^\dag
     L_{P,k} = 0,\\ &L_k=\begin{bmatrix} L_{S,k} & L_{P,k} \\ 0 &
     L_{R,k} \end{bmatrix} \quad \forall k.
  \end{aligned} \right.
\end{equation}
\end{prop}
\noindent

\noindent The following Corollary can be derived in a straightforward
way from Proposition \ref{pro-markovian-invariant-generators}, and it
will be key to establishing the results of the next section:

\begin{cor}[{$R$-Subspace Invariance}] 
\label{upsidedown} 
Consider a QDS on $\Hi=\Hi_S \oplus \Hi_R$, and let the generator
$\Li$ in \eqref{eq:lindblad} be associated to an Hamiltonian $H$ and a
set of noise operators $\{L_k\}$. Then $\Hi_R$ is invariant if and
only if the following conditions hold:
\begin{equation} 
\label{Rinvariance}
  \left \{ \begin{aligned} & iH_P+\frac{1}{2}\sum_k L_{Q,k}^\dag
     L_{R,k} = 0,\\ &L_k=\begin{bmatrix} L_{S,k} & 0 \\ L_{Q,k} &
     L_{R,k} \end{bmatrix} \quad \forall k.
  \end{aligned} \right.
\end{equation}
\end{cor}

One of our aims in this paper is to determine a choice controls that
render an invariant subspace also Globally Asymptotically Stable
(GAS). That is, we wish the target subspace $\Hi_S$ to be both
invariant and {\em attractive}, so that the following property is
obeyed:
$$\lim_{t\rightarrow\infty}\delta({\cal T}_t(\rho),\initset{S}{})=0,
\;\;\;\forall \rho\in\dopset{},$$ 
\noindent 
where $\delta(\sigma,{\cal W}):=\inf_{\tau\in{\cal W}}\|\sigma -
\tau\|$.  In \cite{ticozzi-markovian}, a number of results concerning
the stabilization of pure states and subspaces by both open-loop and
feedback protocols have been established.  For time-independent
Hamiltonian control, in particular, the following condition may be
derived from \eqref{invarianceconditions} above:

\begin{cor}[Open-loop Invariant Subspace]
\label{openloopinvariant} 
Let $\Hi=\Hi_S\oplus\Hi_R.$ Assume that we can modify the system
Hamiltonian as $H'=H+H_c,$ with $H_c$ being an arbitrary,
time-independent control Hamiltonian. Then ${\frak I}_S(\Hi)$ can be
made invariant under ${\cal L}$ if and only if $L_{Q,k}=0$ for every
$k$.
\end{cor}

The following theorems (proved in \cite{ticozzi-markovian}) provide a
general necessary and sufficient condition for attractivity, and
characterize the ability of Hamiltonian control at ensuring the
desired stabilization:

\begin{thm}[{Subspace Attractivity}]
\label{GAS} 
Let $\Hi=\Hi_S\oplus\Hi_R$, and assume that $\Hi_S$ is an invariant
subspace for the QDS dynamics in \eqref{eq:lindblad}. Let \beq
\label{Rprime} \Hi_{R'}= \bigcap_{k=1}^{p} \ker(L_{P,k}), \eeq with
each matrix block $L_{P,k}$ representing a linear operator from
$\Hi_R$ to $\Hi_S.$ Then $\Hi_{S}$ is GAS under ${\cal L}$ if and only
if $\Hi_{R'}$ does not support any invariant subsystem.
\end{thm}

\begin{thm}[Open-loop Subspace Attractivity]
\label{openloopattractivity} Let 
$\Hi =\Hi_S\oplus\Hi_R,$ with $\Hi_S$ supporting an invariant
subsystem. Assume that we can apply arbitrary, time-independent
control Hamiltonians. Then ${\frak I}_S(\Hi)$ can be made GAS under
${\cal L}$ if and only if ${\frak I}_{R}(\Hi)$ is not invariant.
\end{thm}

From a practical standpoint, a main limitation to implementing the
constructive procedure developed in the proof of Theorem
\ref{openloopattractivity} is that the assumption of access to an
arbitrary control Hamiltonian $H_c$ is typically too strong. Thus, we
shall develop (Sec. \ref{constraints}) an approach that allows us to
characterize whether and how a given stabilization task can be
attained with available (in general restricted) time-independent
Hamiltonian controls, as well as the role of the Hamiltonian component
in determining the speed of convergence.

\section{Analysis and synthesis tools}

\subsection{Dissipation-induced decomposition and stability}
\label{sectionDID}

Suppose that we are given a target subspace $\Hi_S \subseteq \Hi$. By
using Propositions \ref{pro-markovian-invariant-generators}, one can
easily check if it is invariant for a {\em given} QDS. If this is the
case, we next exploit the ideas of Theorem \ref{GAS} to devise a
constructive procedure that further determines whether $\Hi_S$ is
attractive or not.  Explicitly, this can be attained by constructing a
special (canonical) Hilbert space decomposition into orthogonal
subspaces, which is induced by the matrices associated to $H,\{L_k\}$
in \eqref{eq:lindblad}.

\noindent \rule{\linewidth}{1pt} \\[-1mm]
{\em Algorithm for Dissipation-Induced Decomposition } \\ [-2.5mm]
\noindent \rule{\linewidth}{1pt}  

Assume $\Hi_S$ to be invariant. Call $\Hi_R^{(0)}:=\Hi_R,$
$\Hi_S^{(0)}:=\Hi_S,$ choose an orthonormal basis for the subspaces
and write the matrices with respect to that basis. Rename the matrix
blocks $H_{S}^{(0)}:= H_{S},$ $H_{P}^{(0)}:= H_{P},$
$H_{R}^{(0)}:=H_{R},$ $L_{S,k}^{(0)}:=L_{S,k},$
$L_{P,k}^{(0)}:=L_{P,k}$, and $L_{R,k}^{(0)}:=L_{R,k}.$ \\

For $j\geq 0$, consider the following iterative procedure:

\begin{enumerate}
\item Compute the matrix blocks $L_{P,k}^{(j)}$ according to the
decomposition ${\Hi}^{(j)}=\Hi_S^{(j)}\oplus\Hi_R^{(j)}.$
\item Define $ \Hi_R^{(j+1)}:=\bigcap_k\ker L_{P,k}^{(j)}.$
\item Consider the following three sub-cases:
\begin{itemize}
\item[a.] If $ \Hi_R^{(j+1)}=\{0\}$,  define $\Hi_T^{(j+1)}:=\Hi^{(j)}_R.$ 
\\ The  iterative procedure is successfully completed.

\item[b.] If $ \Hi_R^{(j+1)}\subsetneq \Hi_R^{(j)},$ define $\Hi_T^{(j+1)}:
=\Hi_R^{(j)}\ominus\Hi_R^{(j+1)}.$ \\

\item[c.] If $ \Hi_R^{(j+1)}= \Hi_R^{(j)}$ (that is, $L_{P,k}^{(j)} =
0 \; \forall k$), define
$$\tilde{\cal L}_P^{(j)}:=-i {H}_P^{(j)}-\frac{1}{2}\sum_k
{L}_{Q,k}^{(j)\dag}L_{R,k}^{(j)}.$$

\begin{itemize}
\item If $\tilde{\cal L}_P^{(j)}\neq0,$ re-define
$\Hi_R^{(j+1)}:=\ker(\tilde{\cal L}^{(j)}_P)$.\\ If $
\Hi_R^{(j+1)}=\{0\}$, define $\Hi_T^{(j+1)}:=\Hi^{(j)}_R$ and the
iterative procedure is successfully completed. Otherwise define
$\Hi_T^{(j+1)}:=\Hi_R^{(j)}\ominus\Hi_R^{(j+1)}$;

\item If $\tilde{\cal L}^{(j)}_P=0,$ then, by Corollary
\ref{upsidedown}, $\Hi_R^{(j)}$ is invariant, and thus (by Theorem
\ref{GAS}) $\Hi_S$ cannot be GAS.
\end{itemize}

\end{itemize}

\item Define $\Hi_S^{(j+1)}:=\Hi_S^{(j)}\oplus \Hi^{(j+1)}_T.$ To
construct a basis for $\Hi_S^{(j+1)},$ append to the {\em already
defined} basis for $\Hi^{(j)}_S$ an orthonormal basis for $
\Hi^{(j+1)}_T.$

\item Increment the counter $j$ and go back to step 1).
\end{enumerate}

\noindent \rule{\linewidth}{1pt} \\

The algorithm ends in a finite number of steps, since at every
iteration it either stops or the dimension of $\Hi^{(j)}_R$ is reduced
by at least one. As anticipated, its main use is as a constructive
procedure to test attractivity of a given subspace $\Hi_S$:

\begin{prop} 
The algorithm is successfully completed if and only if the target
subspace $\Hi_S$ is GAS ($\initset{S}{}$ is GAS).
\end{prop}

\begin{IEEEproof} 
If the algorithm stops because $\tilde{\cal L}_P^{(j)}=0$ for some
$j$, then Corollary \ref{upsidedown} implies that $\Hi_R$ contains an
invariant subspace, hence $\Hi_{S}$ cannot be GAS. On the other hand,
let us assume that the algorithm runs to completion, achieved at
$j\equiv q$. Then we have obtained a decomposition
$\Hi_R=\Hi^{(1)}_T\oplus\Hi^{(2)}_T\oplus\ldots\oplus\Hi^{(q)}_T,$ and
we can prove by (finite) induction that no invariant subspace is
contained in $\Hi_R.$

Let us start from $\Hi^{(q)}_T.$ By definition, since the algorithm is
completed when $\Hi^{(j+1)}_R=\Hi^{(q+1)}_R=\{0\}$, either $\bigcap_k
\ker (L_{P,k}^{(q)}) =\{0\}$ or $\tilde{\cal L}^{(q)}_P$ is
full-rank. In either case, by Theorem \ref{GAS} and Corollary
\ref{upsidedown}, not only does $\Hi^{(q)}_T$ fails to be invariant,
but it cannot contain an invariant subspace.

Now assume (inductive hypothesis) that
$\Hi^{(\ell+1)}_T\oplus\ldots\oplus\Hi_T^{(q)}$, $\ell+1 <q$, does not
contain invariant subspaces, and that (by contradiction)
$\Hi^{(\ell)}_T\oplus\Hi^{(\ell+1)}_T\oplus\ldots\oplus\Hi^{(q)}_T$
does. Then the invariant subspace should be non-orthogonal to
$\Hi^{(\ell)}_T,$ which is, by definition, the orthogonal complement
of either $\bigcap_k \ker (L_{P,k}^{(\ell -1)})$ 
or $\ker (\tilde{\cal L}_P^{(\ell-1)}).$ But then any state $\rho$
with support only on
$\Hi^{(\ell)}_T\oplus\Hi^{(\ell+1)}_T\oplus\ldots\oplus\Hi^{(q)}_T$
and non-trivial support on $\Hi^{(\ell)}_T$ would violate the
invariance conditions and ``exit'' the subspace. Therefore,
$\Hi^{(\ell)}_T\oplus\ldots\oplus\Hi_T^{(q)}$ does not contain
invariant subspaces. By iterating until $\ell=1$, we have that $\Hi_R$
cannot contain invariant subspaces and, by Theorem \ref{GAS}, the
conclusion follows.
\end{IEEEproof}

Formally, the above construction motivates the following:

\begin{defin} 
Let $\initset{S}{}$ be GAS for the dynamics \eqref{eq:lindblad}.  The
Hilbert space decomposition given by 
\beq
\label{DID}
\Hi=\Hi_S\oplus\Hi^{(1)}_T\oplus\Hi^{(2)}_T\ldots\oplus\Hi^{(q)}_T,\eeq
as obtained with the previous algorithm, is called {\em
Dissipation-Induced Decomposition} (DID). Each of the subspaces
$\Hi^{(i)}_T$ in the above direct sum (or possibly a union of them)
will be referred to as a {\em basin}.
\end{defin}

Partitioning each matrix $L_k$ in blocks according to the DID results
in the following canonical structure, where the upper block-diagonal
blocks establish the {\em dissipation-induced} connections between the
different basins $\Hi_T^{(i)}:$
\begin{eqnarray}
L_k=
\left[\begin{array}{c|cccc}  
\cellcolor[gray]{0.85}L_S & \hat L_P^{(0)} & 0 & \cdots &  \\
\hline	0 & \cellcolor[gray]{0.85}L_T^{(1)} & \hat L_P^{(1)} & 0 & \cdots\\
\vdots & L_Q^{(1)} &\cellcolor[gray]{0.85}L_T^{(2)} & \hat L_P^{(1)} & \ddots\\
 &  \vdots & \ddots & \ddots & \ddots\\
\end{array}\right]_k 
\label{matDID}
\end{eqnarray}
\noindent 
The blocks $L_{T,k}^{(i)}$s are the restrictions of $L_k$ to the
subspaces $\Hi_T^{(i)}$.  Since in step 3.b of the DID algorithm
$\Hi_T^{(j)}$ is defined to be in the {\em complement} of $
\Hi_R^{(j+1)}=\bigcap_k\ker L_{P,k}^{(j)}$, at each iteration the only
non-zero parts of the $L_P^{(j)}$ blocks must be in the $(j,j+1)$
block, which we have denoted by $\hat L_{P,k}^{(j)}$ in
\eqref{matDID}. In the upper-triangular part of the matrix, the other
blocks of any row are thus zero by construction. If some $\hat
L_{P,k}^{(j)}=0$ $\forall\ k$, then either the dynamical connection is
established by the Hamiltonian $H,$ through the block $H^{(j)}_P$ (as
checked in step 3.c), or the target subspace is not GAS.

{\em Remark:} The DID in \eqref{DID} is unique, up to an arbitrary
choice of basis in each of the orthogonal components
$\Hi_S,\,\Hi_T^{(i)}$, $i=1,\ldots, q$.  This freedom may be exploited
to further put the matrix blocks in some (problem-dependent)
convenient form.

We illustrate the DID algorithm with an explicit Example, which will
also be further considered in Sec. \ref{application}.

{\em Example 1:} Consider a bipartite quantum system consisting of two
two-level systems (qubits), and on each subsystem choose a basis
$\{\ket{0}_n,\,\ket{1}_n\},$ with $n=1,2$ labeling the qubit. The
standard (computational) basis for the whole system is then given by
$\{\ket{00},\ket{01},\ket{10},\ket{11}\},$ where
$\ket{xy}:=\ket{x}_1\otimes\ket{y}_2$.  As customary, let in addition
$\{ \sigma_a, \, a=x,y,z\}$ denote Pauli pseudo-spin matrices
\cite{sakurai}, with the ``ladder'' operator $\sigma_+ :=(\sigma_x+i
\sigma_y)/2 \equiv |0\rangle\langle 1|$.
Assume that the dynamics is driven by the following QDS generator:
\beq
\label{xsme} 
\dot \rho={\cal L}[\rho]=-i[H,\rho]+L\rho L^\dag -\frac{1}{2}\{L^\dag
L,\,\rho\}, \eeq 
\noindent 
where
\begin{eqnarray}
H&=&(\smallfrac{1}{2} \sigma_z+\sigma_x)\otimes I + I\otimes 
(-\smallfrac{1}{2}\sigma_z + \sigma_x), 
\label{entH}\\
L&=&\sigma_+\otimes I + I\otimes \sigma_+. 
\label{entL}
\end{eqnarray}
It is easy to verify that the (entangled) state
$\rho_d=\ket{\psi_0}\bra{\psi_0},$ with
\[\ket{\psi_0} = \frac{1}{\sqrt{3}}\,(\ket{00}-\ket{01}+\ket{10}),
\] 
is invariant, that is, ${\cal L}[\rho_d]=0$. We can then construct the
DID and verify that such state is also GAS.

By definition, $\Hi_S^{(0)}=\span\{\ket{\psi_0}\},$ and one can write
its orthogonal complement as
$\Hi_R^{(0)}=\span\{\ket{\psi_1},\ket{\psi_2},\ket{\psi_3}\},$ with
an orthonormal basis being given for instance by:
\beq
\ket{\psi_1} = \ket{11},\;\;\;\; \ket{\psi_2} =
\frac{1}{\sqrt{2}}(\ket{01}+\ket{10}),
\label{entB}\eeq
\beq \ket{\psi_3} = -\sqrt{\smallfrac{2}{3}}\Big(\ket{00}
+\smallfrac{1}{2}(\ket{01}-\ket{10})\Big).
\label{entB3} 
\eeq
\noindent 
We begin the iteration with $j=0$ (step 1), having $L_P^{(0)}=[0\;
0.816\; 0].$ We move on (step 2), by defining $ \Hi_R^{(1)}:=\ker
(L_{P}^{(0)})=\span\{\ket{\psi_1},\ket{\psi_3}\}.$ We next get (step
3.b):
\[\Hi_T^{(1)}:=\Hi_R^{(0)}\ominus\Hi_R^{(1)}=\span\{\ket{\psi_2}\},\] 
so that (step 4):
\[\Hi_S^{(1)}=\Hi_S^{(0)}\oplus\Hi_T^{(1)}=\span\{\ket{\psi_0},\ket{\psi_2}\}.\]
We thus set $j=1$ and iterate, obtaining:
\[L_P^{(1)}=\left[\begin{array}{cc} 0 & 0 \\ 1.414 & 0 \end{array}\right],\]
\[\Hi_R^{(2)}=\ker (L_{P}^{(1)})=\span\{\ket{\psi_3}\},\;
\Hi_T^{(2)}=\span\{\ket{\psi_1}\},\]
\[\Hi_S^{(2)}=\span\{\ket{\psi_0},\ket{\psi_2},\ket{\psi_1}\}.\]
In the third iteration, with $j=2,$ we find that $L_P^{(2)}=[0\; 0 \;
0]^T.$ Hence we would have $\Hi_R^{(3)}=\Hi_R^{(2)},$ so we move to
step 3.c. By computing the required matrix blocks we get:
$$\tilde{\cal L}_P^{(2)}=-i {H}_P^{(2)}-\frac{1}{2}\sum_k
{L}_{Q,k}^{(2)\dag}L_{R,k}^{(2)}=-i[0\; -1.732 \; 0]^T.$$ 
\noindent 
Re-defining $\Hi_R^{(3)}:=\ker(\tilde{\cal L}_P)$, we find that
$\Hi_R^{(3)}=\{0\}$, thus $\Hi_T^{(3)}:=\Hi^{(3)}_R$ and the algorithm
is successfully completed. 

Having support on $\Hi_S$ alone, $\rho_d$ is thus GAS, and in the
ordered basis
$\{\ket{\psi_0},\ket{\psi_2},\ket{\psi_1},\ket{\psi_3}\},$ consistent
with the DID
$\Hi_S\oplus\Hi_T^{(1)}\oplus\Hi_T^{(2)}\oplus\Hi_T^{(3)},$ we have
the following matrix representations (cf. Eq. (\ref{matDID})):
\[L=
\left[\begin{array}{c|ccc}  0 & 0.816 & 0 & 0  \\
				   \hline  0 & 0 & 1.414 & 0 \\
					       0 & 0 & 0 & 0\\
					       0 & -1.155 & 0 & 0\\
\end{array}\right],\]
\[H=
\left[\begin{array}{c|ccc}  0 & 0 & 0 & 0  \\
				   \hline  0 & 0 & 1.414 & -1.732
 \\
					       0 & 1.414 & 0 & 0\\
					       0 & -1.732 & 0 & 0\\
\end{array}\right].\]
\noindent 
It is thus evident how the transitions from $\Hi_T^{(1)}$ to $\Hi_S,$
and from $\Hi_T^{(2)}$ to $\Hi_T^{(1)},$ are enacted by the
dissipative part of the generator, whereas only the Hamiltonian is
connecting $\Hi_T^{(3)}$ to $\Hi_T^{(1)},$ destabilizing
$\ket{\psi_3}.$

\subsection{QDS stabilization under control constraints}
\label{constraints}

The DID algorithm can be used as a design tool to determine whether an
admissible control Hamiltonian may achieve stabilization under
constrained capabilities, $(u_1,\ldots,u_\nu)\in{\cal C}\subset
\R^\nu$.  
Assume we are given a target $\Hi_S$, which need not be
invariant or attractive.  We can proceed in two steps.
 
\subsubsection{Imposing invariance} 

Partition $H, L_k$ according to $\Hi=\Hi_S \oplus \Hi_R.$ If $L_{Q,k}
\neq 0$ for some $k$, then $\Hi_S$ is not invariant and it cannot be
made so by Hamiltonian control, hence it cannot be GAS. On the other
hand, if $L_{P,k} = 0$ for all $k$, then $\Hi_S$ cannot be made GAS by
Hamiltonian open-loop control since $\initset{R}{}$ would necessarily
be invariant too (Theorem \ref{openloopattractivity}).

When $L_{Q,k} = 0$ for all $k$ and there exists a $\bar k$ such that
$L_{P,\bar k}\neq 0,$ we need to compute (Proposition
\ref{pro-markovian-invariant-generators})
$$\tilde{\cal L}_P^{}(u)=iH_P(u)-\frac{1}{2}\sum_k L_{S,k}^\dag
L_{P,k}.$$

If $\tilde{\cal L}_P^{}(u)\neq 0$ for all $u\in{\cal C}$, then the
desired subspace cannot be stabilized. Let $ {\cal C}^{(0)}$ be the
set of controls (if any) such that if $\bar u \in {\cal C}^{(0)},$
then $\tilde{\cal L}_P^{}(\bar u) = 0.$

\subsubsection{Exploring the control set for global stabilization}

Having identified a set of control choices that make $\Hi_S$
invariant, we can then use the DID algorithm to check whether they can
also enforce the target subspace to be GAS.

By inspection of the algorithm, the only step in which a different
choice of Hamiltonian may have a role in determining the attractivity
is 3.c. Assume that we fixed a candidate control input $u,$ we are at
iteration $j$ and we stop at 3.c. Assume, in addition, that the last
constrained set of controls we have defined is ${\cal C}^{(\ell)},$
$0\leq\ell<j$ (in case the algorithm has not stopped yet, this is
${\cal C}^{(0)}$). Two possibilities arise:

\begin{itemize}
\item If $\tilde{\cal L}_P^{(j)}\neq0,$ define ${\cal C}^{(j)}$ as the
subset of ${\cal C}^{(\ell)}$ such that if $\bar u \in {\cal
C}^{(j)},$ then it is still true that $\tilde{\cal L}_P(\bar u) \neq
0.$ Pick a choice of $u\in{\cal C}^{(j)},$ and proceed with the
algorithm. Notice that if there exists a control choice $\hat u$ such
that $\tilde{\cal L}_P(\hat u)$ has full rank, we can pick that and
stop the algorithm, having attained the desired stabilization.

\item If $\tilde{\cal L}^{(j)}_P=0,$ the algorithm would stop since
there would be no dynamical link from $\Hi_T^{(j+1)}$ towards
$\Hi_T^{(j)}$, neither enacted by the noise operator nor by the
Hamiltonian.
Hence, we can modify the algorithm as follows. Let us define ${\cal
C}^{(j)}$ as the subset of ${\cal C}^{(\ell)}$ such that if $\bar u
\in {\cal C}^{(j)},$ then $\tilde{\cal L}_P(\bar u) \neq 0.$ If ${\cal
C}^{(j)}$ is empty, no other choice of control could destabilize
$\Hi_R^{(j+1)}$ and $\Hi_S$ cannot be rendered GAS. Otherwise,
redefine ${\cal C}^{(\ell)}:={\cal C}^{(j)},$ pick a choice of
controls in the new ${\cal C}^{(\ell)}$ (for instance at random), and
proceed with the algorithm {\em going back to step $\ell$}.
\end{itemize}

The above procedure either stops with a successful completion of the
algorithm or with an empty ${\cal C}^{(j)}.$ In the first case the
stabilization task has been attained, in the second it has not, and no
admissible control can avoid the existence of invariant states in
$\Hi_R$.

Note that if each ${\cal C}_j$ (thus ${\cal C}$) is finite, for
instance in the presence of quantized control parameters, the
algorithm will clearly stop in a finite number of steps.  More
generally, in the following Proposition we prove that in the common
case of a multi-interval as the set of admissible controls, the design
algorithm works with probability one:
 
\begin{prop} If ${\cal C}$ is a bounded multi-interval of $\R^\nu$, 
the above modified algorithm will end in a finite number of steps with
probability one.
\end{prop}

\begin{IEEEproof} 
The critical point in attaining GAS is finding a set of
control values that ensures {\em invariance} of the desired set when
the the free dynamics would not. In fact, to this end we need to find
a $u \in {\cal C}^{(0)}=\{u \in {\cal C} | \tilde{\cal L}_P^{}(u)=
0\}$. Being ${\cal C}^{(0)}$ the intersection between a multi-interval
and a $(\nu-1)$-dimensional hyperplane in $\R^\nu,$ the set ${\cal
C}^{(0)}$ belongs to a lower-dimensional manifold than ${\cal C}$.

Once invariance has been guaranteed, we are left with the opposite
problem: at each iteration $j,$ we need to avoid the set of controls
such that $\tilde{\cal L}^{(j)}_P=0.$ This is again a
$(\nu-1)$-dimensional hyperplane in $\R^\nu.$ Therefore, if a certain
$u_0$ is such that $\tilde{\cal L}^{(j)}_P=0$ but not all of them are,
this belongs to a lower-dimensional manifold with respect to ${\cal
C}^{(0)}$.  Hence, picking a random choice of $u\in{\cal C}^{(0)}$
(with respect to a uniform distribution) will almost always guarantee
that the algorithm will stop in a finite number of steps.
\end{IEEEproof}

\subsection{Approximate state stabilization}
\label{practical}

A necessary and sufficient condition for a state (not necessarily
pure) to be GAS is that it is the unique stationary state for the
dynamics \cite{schirmer-markovian}: this fact can be exploited, under
appropriate assumptions, to {\em approximately} stabilize a desired
pure state $\rho_d$ when {\em exact} stabilization cannot be achieved.

Assume that at the first step in the previous procedure we see that
$\rho_d$ is not invariant, even if $L_{Q,k}^{(0)}=0$ for all $k$,
since
$$\tilde{\cal L}_P^{(0)}=i {H}_P^{(0)}-\frac{1}{2}\sum_k
{L}_{S,k}^{(0)\dag}L_{P,k}^{(0)}\neq 0,$$
\noindent 
and there exists no choice of controls that achieve stabilization.  If
however the (operator) norm of $\tilde{\cal L}_P^{(0)}$ can be made
small, in a suitable sense, we can still hope that a GAS state close
to $\rho_d$ exists. This can checked as it follows:

\begin{itemize}
\item Define $\tilde H_P:=i\tilde{\cal L}_P^{(0)}.$ Consider a new
Hamiltonian $$\tilde H:= H^{(0)}+\Delta H= \begin{bmatrix} H_S & H_P
\\ H_P & H_R \end{bmatrix} + \begin{bmatrix} 0 & \tilde H_P \\
\tilde{H}_P^\dag & 0 
\end{bmatrix}.$$
By construction, $\rho_d$ is invariant under $\tilde H$.

\item Proceed with the algorithm described in the previous subsection
in order to stabilize $\rho_d$ with $\tilde H$ instead of $H$.

\item As a by-product, the subset of control values that achieve
stabilization is found. Let it be denoted by ${\cal S}\subseteq {\cal
C}$;

\item Determine $u_* \in {\cal S}$ such that $\min_{u\in{\cal S}}
\|\tilde {\cal L}_P^{(0)}(u)\|_\infty$ is attained.
\end{itemize}

After the control synthesis, the generator for the actual system is in
the form:

$$ \dot{\rho} = \tilde{{\cal L}}[\rho] - \Delta{\cal L}[\rho],$$
\noindent
with $\Delta{\cal L}[\rho]= -i[\Delta H, \rho],$ and $\tilde{{\cal
L}}[\rho]$ having $\rho_d$ as its unique stationary state. It can be
easily proved (by using the standard {\em coherence-vector}
representation, following the same argument of Section III.E in
\cite{ticozzi-QDS}) that the perturbed generator will still have a
unique stationary state provided that the norm of $\Delta{\cal L}$ is
small, with respect to the smallest absolute value of the non-zero
eigenvalues of $\tilde{{\cal L}}$.  In our setting, $\|\Delta{\cal
L}\|$ can be bounded by twice the (operator) norm of $\tilde H_P$. By
continuity arguments, one can also show that the new stationary state
will then be close to the target one.  This ensures stabilization of a
(generally) mixed state in a neighborhood of $\rho_d$ or, in the
control-theoretic jargon, ``practical stabilization'' of the target
state, the size of the neighborhood depending on $\|\Delta{\cal L}\|$.
 
\section{Speed of convergence of a QDS}

{\em How quickly can the system reach the GAS subspace $\Hi_S$ from a
generic initial state?}  We address this question in two different
ways. The first approach relies on explicitly computing the asymptotic
speed of convergence by considering the spectrum of ${\cal L}$
(sp$({\cal L})$ henceforth) as a linear superoperator. Despite its
simplicity and natural system-theoretic interpretation, the resulting
(worst-case) bound provides little (if any) physical intuition on what
effect individual control parameters have on the overall dynamics.  In
the second approach, which builds directly on the DID, we argue that
convergence is limited by the slowest speed of transfer from a basin
subspace to the preceding one in the chain.  Despite its qualitative
nature, this has the advantage of offering a more transparent physical
picture and eventually some useful criteria for the design of rapidly
convergent dynamics.

\subsection{System-theoretic approach}
\label{speednum}

The basic step is to employ a vectorized form of the QDS generator
${\cal L}$ (also known as superoperator or Liouville space formalism
in the literature \cite{nielsen-chuang}), in such a way that standard
results on linear time-invariant (LTI) state-space models may be
invoked.  Recall that the vectorization of a $n\times m$ matrix $M$,
denoted by $\vec(M)$, is obtained by stacking vertically the $m$
columns of $M$, resulting in a $n^2$-dimensional vector
\cite{horn-johnson}.  Vectorization is a powerful tool when used to
express matrix multiplications as linear transformations acting on
vectors.  The key property we will need is the following: For any
matrices $X$, $Y$ and $Z$ such that their composition $XYZ$ is well
defined, it holds that:
\begin{equation} 
\label{eq-property-vectorization}
\vec(XYZ) = (Z^T \otimes X)\vec(Y), 
\end{equation}
where the symbol $\otimes$ is to be understood here as the Kronecker
product of matrices.
The following Theorem provides a necessary and sufficient condition
for GAS subspaces directly in terms of spectral properties of the
(vectorized) generator (compare with Theorem \ref{GAS}):

\begin{thm}[Subspace Attractivity] 
\label{thm-attracting-subspace-vec}
Let $\Hi=\Hi_S\oplus\Hi_R$, and assume that $\Hi_S$ is an invariant
subspace for the QDS dynamics in \eqref{eq:lindblad}.  Then $\Hi_S$ is
GAS if and only if the linear operator defined by the equation
\begin{eqnarray} 
    \hatLi_R := &\hspace*{-2mm}-\hspace*{-2mm}& \frac {i}{\hbar} 
\left(\identity_R \otimes H_R -
H_R^T \otimes \identity_R \right)
+\sum_k L_{R,k}^* \otimes L_{R,k} \nonumber \\ 
&\hspace*{-2mm}- \hspace*{-2mm}& \frac{1}{2} \sum_k 
\identity_R \otimes (L_{P,k}^\dag
L_{P,k} + L_{R,k}^\dag L_{R,k})\label{eq-hatL-R} \\ 
& \hspace*{-2mm}-\hspace*{-2mm}&\frac{1}{2} \sum_k (L_{P,k}^T L_{P,k}^* 
+ L_{R,k}^T L_{R,k}^*)\otimes \identity_R. \nonumber
\end{eqnarray}
does \emph{not} have a zero eigenvalue.
\end{thm}

\begin{IEEEproof}
Let $\bar{\Pi}_R = \begin{bmatrix}0 &\identity_R\end{bmatrix}$. By
explicitly computing the generator's $R$-block we find:
\begin{equation}\label{eq-rho_R-generator}
  \begin{aligned}
    \bar{\Pi}_R \Li[\rho]\bar{\Pi}_R^\dag = & -\frac
                 {i}{\hbar}\big([H_R, \rho_R] + H_P^\dag \rho_P
                 -\rho_P^\dag H_P \big) \\ &\;\; +\sum_k
                 \big(L_{Q,k}\rho_S +
                 L_{R,k}\rho_P^\dag\big)L_{Q,k}^\dag\\ &\;\; +\sum_k
                 \big(L_{Q,k}\rho_P + L_{R,k}\rho_R\big)
                 L_{R,k}^\dag\\ &\;\; -\frac{1}{2} \sum_k
                 \big(L_{P,k}^\dag L_{S,k} + L_{R,k}
                 L_{Q,k}\big)\rho_P\\ &\;\; -\frac{1}{2} \sum_k
                 \rho_P^\dag \big(L_{S,k}^\dag L_{P,k} + L_{Q,k}^\dag
                 L_{R,k}\big) \\ &\;\; -\frac{1}{2} \sum_k
                 \big\{L_{P,k}^\dag L_{P,k} + L_{R,k}^\dag
                 L_{R,k},\rho_R\big\}.
  \end{aligned}
\end{equation}
Since $\Hi_S$ is invariant, the evolution of the state's $R$-block
turns out to be decoupled from the rest: in fact, by substituting the
invariance conditions \eqref{invarianceconditions} into
\eqref{eq-rho_R-generator}, we find:
\begin{equation}
  \begin{aligned} 
   \label{eq-rho_R-decoupled-generator}
    \bar{\Pi}_R \Li[\rho]\bar{\Pi}_R &= -\frac {i}{\hbar}[H_R, \rho_R]
             +\sum_k L_{R,k}\rho_R L_{R,k}^\dag\\ &\quad
             -\frac{1}{2}\sum_k \big\{L_{P,k}^\dag L_{P,k} +
             L_{R,k}^\dag L_{R,k},\rho_R\big\}.
  \end{aligned}
\end{equation}
Now let $\hat\rho_R = \vec(\bar{\Pi}_R \rho \bar{\Pi}_R^\dag)$. By
exploiting the composition property \eqref{eq-property-vectorization}, 
we have:
$$ \dot{\hat{\rho}}_R = \hatLi_R \hat\rho_R,$$ 
\noindent 
where $\hatLi_R$ is the map defined in \eqref{eq-hatL-R}.

Suppose that $\Hi_S$ is \emph{not} attractive. By Theorem \ref{GAS},
the dynamics must then admit an invariant state with support on
$\Hi_R$.  This implies that $\hat \Li_R$ has at least one non-trivial
steady state, corresponding to a zero eigenvalue.  To prove the
converse, suppose that $(0,\vec(X))$ is an eigenpair of
$\hat\Li_R$. Clearly, $X\neq0$ by definition of eigenvector. Then, any
initial state $\rho \in \dopset{}$ such that its $R$-block, $\rho_R$,
has non-vanishing projection on $X$ cannot converge to
$\initset{S}{},$ and thus $\Hi_S$ is not attractive. Since $\dopset{}$
contains a set of generators for $\bounded{}$ (e.g. the pure states),
there is at least one state such that $\tr(\rho X)\neq 0$.
\end{IEEEproof}

Building on Theorem \ref{thm-attracting-subspace-vec}, the following
Corollary gives a bound on the asymptotic convergence speed to an
attractive subspace, based on the modal analysis of LTI systems:

\begin{cor}[Asymptotic convergence speed]
\label{corollary-Z-asymp-bound}
Consider a QDS on $\Hi = \Hi_S \oplus \Hi_R$, and let $\Hi_S$ be a GAS
subspace for the given QDS generator. Then any state $\rho \in
\dopset{}$ converges asymptotically to a state with support only on
$\Hi_S$ at least as fast as $ke^{\lambda_0 t}$, where $k$ is a
constant depending on the initial condition and $\lambda_0$ is given
by:
\begin{equation}\label{speed}
  \lambda_0 = \max_\lambda \{\real(\lambda)\;|\; \lambda \in
  \spec(\hatLi_R)\}.
\end{equation}
\end{cor}

\noindent Note that, in the case of one-dimensional $\Hi_S$, the
``slowest'' eigenvalue $\lambda_0$ is also the smallest Lyapunov
exponent of the dynamical system \eqref{eq:lindblad}.

\subsection{Connected basins approach}
\label{basins}

Recall that the DID derived in Section \ref{sectionDID} is a
decomposition of the systems's Hilbert space in orthogonal subspaces:
$$\Hi=\Hi_S\oplus\Hi^{(1)}_T\oplus\Hi^{(2)}_T\ldots\oplus\Hi^{(q)}_T.$$
\noindent 
By looking at the block structure of the matrices $H,L_k$ induced by
the DID, we can classify each basin depending on how it is {\em
dynamically connected} to the preceding one in the DID. Beside
$\Hi_S,$ which is assumed to be globally attractive and we term the
{\em collector basin}, let us consider a basin $\Hi_B=\Hi_T^{(i)},$
or, more generally,
$\Hi_B=\Hi^{(i)}_T\oplus\ldots\oplus\Hi^{(i+q)}_T.$ We can distinguish
the following three possibilities for $\Hi_B$ :

\begin{enumerate}

\item[A.]{\em Transition basin}: This allows a one-way connection from
$\Hi_T^{(i)}$ to $\Hi_T^{(i-1)},$ when the following conditions are
satisfied:
$$ \hat L^{(i-1)}_{P,k}\neq 0 \textrm{ for some } k,\quad
L^{(i-1)}_{Q,k} = 0\; \forall\,k,$$ in addition to the invariance
condition \beq\label{condinv}i {H}_P^{(i-1)}-\frac{1}{2}\sum_k
{L}_{S,k}^{(i-1)\dag}L_{P,k}^{(i-1)}=0.\eeq In other words, $\hat
L_{P,k}^{(i-1)}$ enacts a directed {\em probability flow} towards the
beginning of the DID: states with support on $\Hi_T^{(i)}$ are
``pulled'' towards $\Hi_T^{(i-1)}.$

\item[B.] {\em Mixing basin}: This allows for the dynamical connection
between the subspaces to be bi-directional, which occurs in the
following cases, or {\em types}:
\begin{enumerate}
\item[1.] As in the transition basin, but with $$i
{H}_P^{(i-1)}-\frac{1}{2}\sum_k
{L}_{S,k}^{(i-1)\dag}L_{P,k}^{(i-1)}\neq 0;$$
\item [2.]  In the generic case, when both $ \hat L^{(i-1)}_{P,k}\neq
0$, $L^{(i-1)}_{Q,k'}\neq 0 \textrm{ for some } k,k';$
\item[3.] When $ \hat L^{(i-1)}_{P,k}= 0\; \forall\,k,\;
L^{(i-1)}_{Q,k} \neq 0,\textrm{ for some } k.$
\end{enumerate}

\item[C.]{\em Circulation basin}: In this case, $ \hat
L^{(i-1)}_{P,k}= 0$ and $L^{(i-1)}_{Q,k} = 0$ for all $k,$ and thus
the dynamical connection is enacted solely by the Hamiltonian block
$H_P$.  Not only is the dynamical connection bi-directional, but it is
also ``symmetric'' and, {\em in the absence} of dynamics internal to the
subspaces $\Hi_T^{(i)},\Hi_T^{(i-1)},$ and further connections, the
state would keep ``circulating'' between the subspaces.
\end{enumerate}
   
How is this related to the speed of convergence? Let us consider a
pair of basins $\Hi_T^{(i-1)},$ $\Hi_T^{(i)},$ and let us try to
investigate how fast a state with support only in $\Hi_T^{(i)}$
can ``flow'' towards $\Hi_T^{(i-1)}$ in a worst case scenario.
The answer depends on the dynamical connections, that is, the kind of
basin the state is in, and the required speed can in each case be
estimated as follows:

\begin{itemize}

\item[(i)] {\em Transition basin, type-1 and type-2 mixing basins}:
The exact speed of ``exit'' from a transition basin $\Hi_T^{(i)}$ has
been calculated in \cite{ticozzi-markovian}, and reads
\beq\label{speed1}\lambda_i(\rho)=\tr\Big(\sum_k \hat L_{P,k}^{(i-1)\dag} \hat
L_{P,k}^{(i-1)}\rho\Big),\eeq 
\noindent 
which in the worst case scenario corresponds to {\em the minimum
eigenvalue of $ \sum_k \hat L_{P,k}^{(i-1)\dag} \hat
L_{P,k}^{(i-1)}$},
\[\hat\gamma_i^L=\min\Big\{\lambda|\lambda\in\spec\Big(\sum_k 
\hat L_{P,k}^{(i-1)\dag} \hat L_{P,k}^{(i-1)}\Big)\Big\}.\] 
\noindent 
The same quantity works as an estimate for the {\em mixing basin of
type-1 and-2,} since to first order, with a state with support on
$\Hi_T^{(i)}$ alone, the invariance condition \eqref{condinv} and the
$L^{(i-1)}_{Q,k}$ blocks have no influence.  Let us recall that in
order to have a GAS subspace, at least $\Hi_T^{(1)}$ has to be a {\em
transition} basin.

\item[(ii)] {\em Mixing basin of type-3 and circulation basin}: When
$\hat L^{(i-1)}_{P,k}= 0$ for all $k$, and $L^{(i-1)}_{Q,k} \neq
0,\textrm{ for some } k,$ the exit from $\Hi_T^{(i)}$ is determined by
the Hamiltonian, and hence it is a second order effect. Therefore, it
is not possible to estimate the ``transfer speed'' based on the
derivative of the trace as we did above, since the latter would be
zero. Similarly, when the dynamical connection is enacted only by the
Hamiltonian block $H_P$, then again, it is a second order effect.

Let us restrict our attention to the relevant subspace,
$\Hi_{(i-1,i)}=\Hi_T^{(i-1)}\oplus\Hi_T^{(i)},$ and write the
restricted Hamiltonian in block-form:
\[\Pi_{\Hi_{(i-1,i)}}H\Pi_{\Hi_{(i-1,i)}}=
\left[\begin{array}{c|c}
 H_T^{(i-1)} & H_P^{(i-1)}\\ \hline
 H_P^{(i-1)\dag} & H_T^{(i) }
\end{array}\right].\]

We can always find a unitary change of basis $U_T^{(i-1)}\oplus
U_T^{(i)}$ that preserves the DID and it is such that
$U_T^{(i-1)}H_P^{(i-1)}U_T^{(i)\dag}=\Sigma_P^{(i-1)}>0,$ with
$\Sigma_P^{(i-1)}=\diag(s_1,\ldots,s_{d_i})$ being the diagonal matrix
of the singular values of $H_P^{(i-1)}$ in decreasing order.  Then the
effect of the off-diagonal blocks is to couple pairs of the new basis
vectors in $\Hi_T^{(i-1)},\Hi_T^{(i)}$ generating simple rotations of
the form $e^{-is_j\sigma_xt}$.  Hence, any state in $\Hi_T^{(i)}$ will
rotate towards $\Hi_T^{(i-1)}$ as a (generally time-varying, due to
the diagonal blocks of the Hamiltonian) combination of cosines,
$\sum_k\ell_k(t)\cos(s_kt)$, for appropriate coefficients. The
required estimate in this case can be obtained by comparing the speed
of transfer induced by the Hamiltonian coupling to the exponential
decay in the previous case. When the noise action is dominant,
$\hat\gamma^L_i$ can be thought as $1/\hat T,$ with $\hat T$ being the
``decoherence time scale'' needed for the associated decaying
exponential to reach the value $e^{-1}$. Comparing with the action of
$H_P^{(i)}$ by letting $\hat s_i=\min_j\{s_j\}$ then yields:
$$\hat\gamma_i^H= \hat s_i/\arccos(e^{-1}).$$ This formula does not
takes into account the effect of the diagonal block of the
Hamiltonian, whose influence and optimization will be studied in
Subsection \ref{mixham}.
\end{itemize}

It is worth remarking that the ``transfer'' is monotone in case (i),
whereas in case (ii) it is so only in an initial time interval.  Once
we establish an estimate for all the transition speeds, we can think
of {\em the slowest speed,} call it $\gamma_{\text{min}},$ as the
``bottleneck'' in order to attain fast convergence.  If, in
particular, $\gamma_{\text{min}}=\hat\gamma_i^L$ for a certain $i$,
since the latter is not affected by the Hamiltonian, it provides a
fundamental limit to the attainable convergence speed given purely
Hamiltonian control resources. In fact, imagine the system to be
prepared in a state of $\Hi_T^{(q)},$ the last basin in the DID, and
follow its flow towards the attractive collector basin $\Hi_S.$ In the
worst case, it will have to move through the slowest possible
connection, associated to $\gamma_{\text{min}}$.  Conversely,
connections enacted by the Hamiltonian can in principle be optimized,
following the design prescriptions we shall outline below.

In spite of giving only qualitative indications, the advantage of this
approach is twofold: (i) Estimating the transition speed between
basins is, in most practical situations, both simpler and more
efficient than deriving closed-form expressions for the eigenvalues of
the generator; (ii) Unlike the system-theoretic approach, it gives an
immediate idea of which control parameters have a role in influencing
the speed of convergence.


\subsection{Tuning the convergence speed via Hamiltonian control}
\label{mixham}

It is well known that the interplay between dissipative and
Hamiltonian dynamics is critical for controllability
\cite{altafini-open}, invariance, asymptotic stability and
noiselessness \cite{ticozzi-QDS,ticozzi-markovian}, as well as for
purity dynamics \cite{bartana-tannor}.  By recalling the definition of
$\hat {\cal L}_R$ given in Eq. \eqref{eq-hatL-R}, Corollary
\ref{corollary-Z-asymp-bound} highlights that not only can the
Hamiltonian have a key role in determining the stability of a given
state, but it can also influence significantly the convergence speed.
In order to gain concrete insight, we first study a simple
prototypical example.

{\em Example 2:} Consider a three-dimensional system driven by a
generator of the form \eqref{eq:lindblad}, with operators $H,L$ that
with respect to the (unique, in this case) DID basis
$\{\ket{s},\ket{r_1},\ket{r_2}\}$ have the following form:
\begin{equation} 
\label{eq-qds3d-matrices}
  H = \left[ \begin{array}{c|c c}
               \Upsilon & 0 & 0\\ \hline
               0 & \Delta & \Omega\\
               0 & \Omega & 0 \\                            
             \end{array}\right ] , \quad
  L = \left[ \begin{array}{c|c c}
               0 & \ell & 0\\ \hline
               0 &  0   & 0\\
               0 &  0   & 0 \\                            
             \end{array}\right ].
\end{equation}
\noindent 
It is easy to show, by recalling Proposition 1, that
$\rho_d=\ket{s}\bra{s}$ is invariant, and that any choice of
$\Omega,\ell\neq 0$ also renders $\rho_d$ GAS.  We can study the
eigenvalues of $\hat {\cal L}_R$ as defined in \eqref{eq-hatL-R} and
invoke Theorem \ref{thm-attracting-subspace-vec}.

\begin{figure}[t]
  \centering
  \includegraphics[width=7cm]{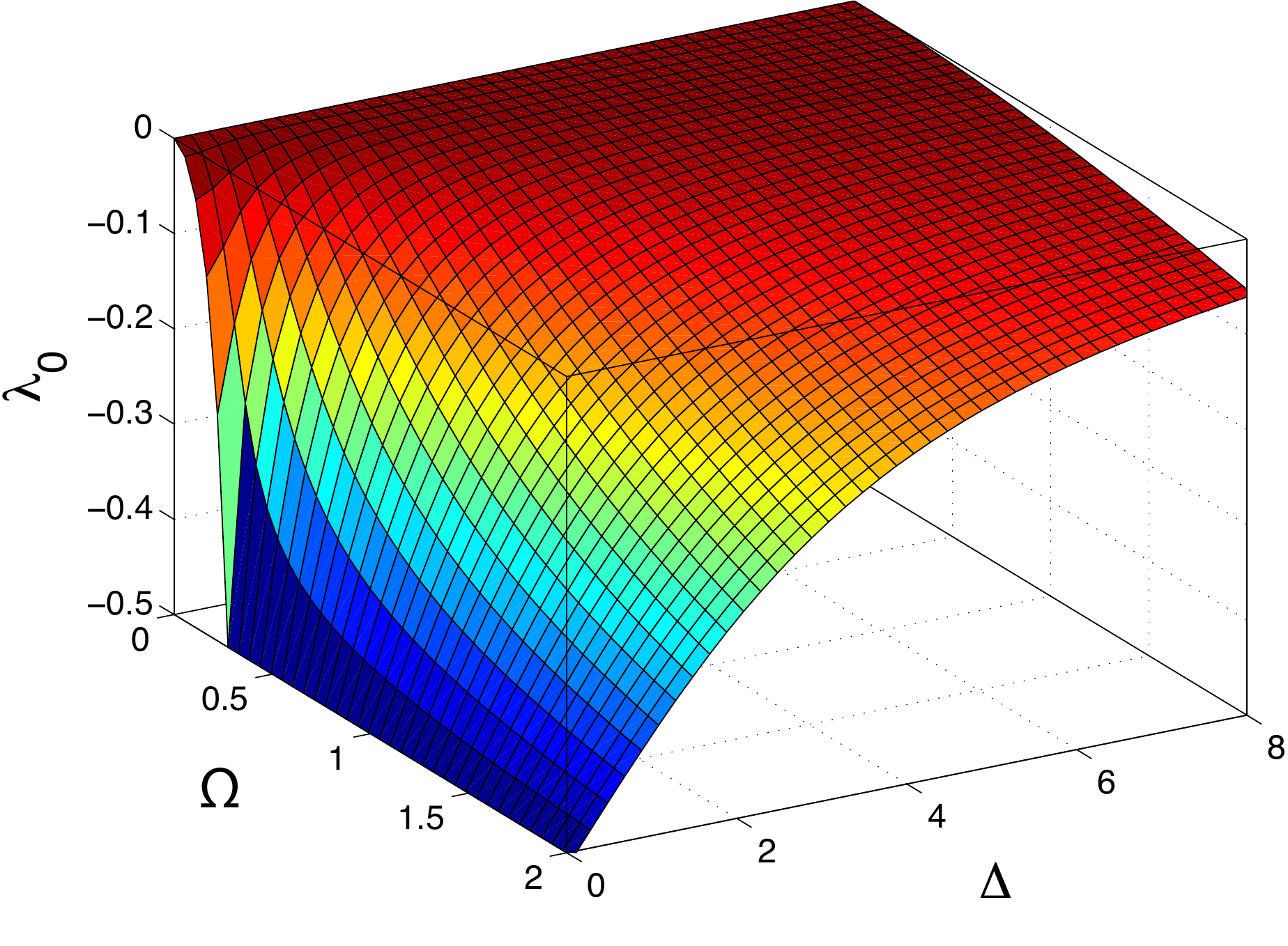}
  \caption{Convergence speed to the GAS target state $\rho_d =
    \ket{s}\bra{s}$ for the 3-level QDS in Example $2$ as a function
    of the time-independent Hamiltonian control parameters $\Delta$
    and $\Omega$. }
\label{fig1}
\end{figure}

Without loss of generality, let us set $\ell=1$ so that
$L=|s\rangle\langle r_1|$, and assume that $\Delta,\ \Omega$ are
positive real numbers, which makes all the relevant matrices to be
real.  Let $\Pi_0:=L_P^T L^P$. We can then rewrite \beq
\label{reduct}
\hat{\cal L}_R= R^+\otimes I_R + I_R\otimes R^-,\eeq
\noindent 
where $R^\pm=\pm iH_R-\Pi_0/2$. Let $\lambda_{1,2}^\pm$ be the
eigenvalues of $R^+, R^-$.  Given the tensor structure of $\hat{\cal
L}_R,$ the eigenvalues of \eqref{reduct} are simply
$\alpha_{ij}=\lambda_{i}^++\lambda_{j}^-,$ with $i,j=1,2$.  The real
parts of the $\alpha_{ij}$ can be explicitly computed:
$$\Re(\alpha_{ij})=[-1/2,-1/2,-1/2\pm1/2\real(\sqrt{\Gamma})],$$ where
$\Gamma=1-\Delta^2+i\Delta-4 \Omega ^2.$ The behavior of
$\lambda_0=-1/2+1/2\Re(\sqrt{\Gamma})$ is depicted in Figure
\ref{fig1}. Two features are apparent: Higher values of $\Omega$ lead
to faster convergence, whereas higher values of $\Delta$ slow down
convergence.  The optimal scenario ($|\lambda_0|=1/2$) is attained for
$\Delta=0$.

The above observations are instances of a general behavior of the
asymptotic convergence speed, when the Hamiltonian provides a
``critical" dynamical connection between two subspaces.  Specifically,
the off-diagonal part of $H_R$ in \eqref{eq-qds3d-matrices} is
necessary to make $\rho_d$ GAS, connecting the basins associated to
$\ket{r_1}$ and $\ket{r_2}$. Nonetheless, the {\em diagonal} elements
of $H$ also have a key role: their value influences the positioning of
the energy eigenvectors, which by definition are the system states
that are not affected by the Hamiltonian action. Intuitively, under
the action of $H$ alone, all other states ``precess" unitarily around
the energy eigenvectors, hence the closer the eigenvectors of $H$ are
to the the states we aim to {\em destabilize}, the weaker the
destabilizing action will be.

A way to make this picture more precise is to recall that
$\rho_d=\ket{s}\bra{s}$ is invariant, and that the basin associated to
$\ket{r_1}$ is directly connected to $\rho_d$ by dissipation.  Thus,
in order to make $\rho_d$ GAS we only need to destabilize $\ket{r_2}$
using $H$.  Consider the action of $H$ restricted to
$\Hi_R=\span(\ket{r_1},\ket{r_2})$.

The Hamiltonian's $R$-block spectrum is given by
\begin{equation}
\label{spectrumR}
  \spec(H_R) = \Big\{\frac{\Delta \pm \sqrt{\Delta^2 + 4 \Omega
  ^2}}{2}\Big\},
\end{equation}
with the correspondent eigenvectors:
\begin{equation}
\label{eigvecR}
  \ket{\pm} = \smallfrac{2 \Omega}{\sqrt{8 \Omega ^2+2\Delta^2\pm
  2\Delta\sqrt{\Delta^2 + 4 \Omega ^2}}}\begin{bmatrix} 1\\
  -\frac{\Delta \pm\sqrt{\Delta^2 + 4 \Omega ^2}}{2 \Omega}
  \end{bmatrix}.
\end{equation}
\noindent 
Decreasing $\Delta,$ the eigenvectors of $H_R$ tend to
$(\ket{r_1}\pm\ket{r_2})/\sqrt{2}$, and if $\Delta=0$, the state
$\ket{r_2},$ which is unaffected by the noise action, is rotated right
into $\ket{r_1}$ after half-cycle.  Physically, this behavior simply
follows from mapping the dynamics within the $R$-block to a Rabi
problem (in the appropriate rotating frame), the condition $\Delta=0$
corresponding to resonant driving \cite{sakurai}.

Beyond the specific example, our analysis suggests two guiding
principles for enhancing the speed of convergence via
(time-independent) Hamiltonian design.  Specifically, one can: 
\begin{itemize}
\item Augment the dynamical connection induced by the Hamiltonian by
larger off-diagonal couplings;
\item Position the eigenvectors of the Hamiltonian as close as
possible to balanced superposition of the state(s) to be destabilized
and the target one(s).
\end{itemize}


\section{Applications}
\label{application}

In this Section, we analyze three examples that are directly inspired
by physical applications, with the goal of demonstrating how the
control-theoretic tools and principles developed thus far are useful
to tackle stabilization problems in realistic quantum-engineering
settings.  

\subsection{Attractive decoherence-free subspace in an optical system}

\begin{figure}
  \centering \includegraphics[width=5.3cm]{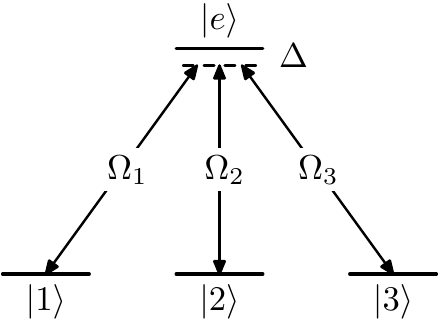}
  \caption{Energy level configuration of the 4-level optical system
    discussed in Section V.A. Three degenerate stable states are
    coupled to an excited state trough separate laser fields with a
    common detuning $\Delta$ and amplitude $\Omega_i$.}
    \label{fig-ex-real-atomic-conf}
\end{figure}

Consider first the quantum-optical setting investigated in
\cite{YHWO09}, where Lyapunov control is exploited in order to drive a
dissipative four-level system into a decoherence free subspace (DFS).
A schematic representation of the relevant QDS dynamics is depicted in
Figure \ref{fig-ex-real-atomic-conf}. Three (degenerate) stable ground
states, $\ket{i}_{i=1,2, 3}$, are coupled to an unstable excited state
$\ket{e}$ through three separate laser fields characterized by the
coupling constants $\Omega_i$, $i=1,2,3$. In a frame that rotates with
the (common) laser frequency, the Hamiltonian reads 
\begin{equation}
H = \Delta \ket{e}\bra{e} + \sum_{i=1}^3 \Omega_i\big(\ket{e}\bra{i} +
\ket{i}\bra{e}\big),
\end{equation}
where $\Delta$ denotes the detuning from resonance.  The decay of the
excited state to the stable states is a Markovian process
characterized by decay rates $\gamma_i$, $i=1,2,3$. The relevant
Lindblad operators are thus given by the atomic lowering operators
$L_i = \sqrt{\gamma_i}\,\ket{i}\bra{e}$, $i=1,2,3$.

Let the coupling factors $\Omega_i$ be parameterized as 
\begin{equation} 
\label{eq-ex-real-spherical-coords}
  \left \{ \begin{aligned}
    \Omega_1 &= \Omega \sin\theta  \cos\phi ,\\
    \Omega_2 &= \Omega  \sin\theta  \sin\phi ,\\
    \Omega_3 &= \Omega  \cos\theta ,
  \end{aligned} \right.
\end{equation}
where $\Omega \geq 0$ and $0\leq \theta \leq \pi$, $0\leq \phi <
2\pi$, respectively.  It is known (see e.g. \cite{YHWO09} and
references therein) that the resulting generator admits a DFS
$\Hi_{\text{DFS}}$ spanned by the following orthonormal basis:
\begin{equation} 
\label{eq-dfs-basis}
  \left \{ \begin{aligned} \ket{d_1} &= - \sin \phi \ket{1} + \cos
    \phi \ket{2}, \\ \ket{d_2} &= \cos \theta (\cos \phi\ket{1} + \sin
    \phi \ket{2}) -\sin \theta \ket{3},
  \end{aligned} \right.
\end{equation}
provided that $\theta \neq k\pi$ and $\phi \neq \frac{\pi}{2}
+k\pi$. In order to formally establish that this DFS is also GAS for
almost all choices of the QDS parameters $\Delta,\Omega_i,\gamma_i$,
we construct the DID starting from $\Hi_S=\Hi_{\text{DFS}}$, and
obtaining $\Hi_T^{(1)}= \textrm{span}\{\ket{e}\},$
$\Hi_T^{(2)}=\Hi\ominus (\Hi_{\text{DFS}}\oplus\Hi_T^{(1)})$.  The
corresponding matrix representation of the Hamiltonian and noise
operators becomes:
\begin{eqnarray*} 
\label{eq-qds4d-matrices}
  \begin{aligned}
    H &= \left[ \begin{array}{cc|c|c}
                  0 & 0 & 0 & 0 \\
                  0 & 0 & 0 & 0 \\ \hline
                  0 & 0 & \Delta & \Omega' \\\hline
                  0 & 0 & \Omega' & 0 \\ 
                \end{array}\right ] ,\\
    L_1 &= \left[ \begin{array}{cc|c|c}
                    0 & 0 & -\sqrt{\gamma_1}\,\sin\phi & 0 \\ 
                    0 & 0 & \sqrt{\gamma_1}\,\cos\phi \,\cos\theta & 0  \\ \hline
                    0 & 0 & 0 & 0  \\ \hline
                    0 & 0 & \sqrt{\gamma_1} |\cos\phi\, \sin\theta| & 0
                  \end{array}\right ], \\
    L_2 &= \left[ \begin{array}{cc|c|c}
                    0 & 0 & \sqrt{\gamma_2}\,\cos\phi & 0 \\ 
                    0 & 0 & \sqrt{\gamma_2}\,\cos\theta \, \sin\phi & 0 \\ \hline
                    0 & 0 & 0 & 0 \\ \hline
                    0 & 0 & \sqrt{\gamma_2} \,\sign(\cos\phi)\sin\phi|\sin\theta| & 0
                  \end{array}\right ], \\
    L_3 &= \left[ \begin{array}{cc|c|c}
                    0 & 0 & 0 & 0 \\ 
                    0 & 0 & -\sqrt{\gamma_3}\,\sin\theta & 0 \\ \hline
                    0 & 0 & 0 & 0 \\ \hline
                    0 & 0 & \sqrt{\gamma_3}\,\sign(\cos\phi\, \sin\theta)\cos\theta & 0 
                  \end{array}\right ],\\
   \end{aligned}
\end{eqnarray*}
where $\sign(x)$ is the sign function and $\Omega' := \Omega\,
\sign(\sin\theta \cos\phi)$.  By Proposition
\ref{pro-markovian-invariant-generators}, it follows that
$\Hi_{\text{DFS}}$ is invariant. Furthermore, the vectorized map
governing the evolution of the state's $R$-block in \eqref{eq-hatL-R}
has the form:
\begin{equation*}
  \hatLi_R =\begin{bmatrix} -\sum_i \gamma_i & -i\Omega'  & i\Omega' 
  & 0\\ -i\Omega' & -\sum_i \frac{\gamma_i}{2} + i\Delta  & 0 &
  i\Omega' \\ i\Omega' & 0 & - \sum_i \frac{\gamma_i}{2} - i\Delta  &
  -i\Omega'\\ \sum_i \gamma_i \frac{\Omega_i^2}{\Omega^2} & i\Omega' 
  & -i\Omega'  & 0 \end{bmatrix}.
\end{equation*}
Then, by Theorem \ref{thm-attracting-subspace-vec}, a sufficient and
necessary condition for $\Hi_{\text{DFS}}$ to be GAS is that the
characteristic polynomial of $\hatLi_R$, $\Delta_{\hatLi_R}(s)$, has
no zero root. Explicit computation yields:
\begin{equation}
\Delta_{\hatLi_R}(0) = \Big(\sum_i \gamma_i\Big)\Big( \sum_i
\gamma_i(\Omega^2 - \Omega_i^2)\Big),
\end{equation}
which clearly vanishes in the trivial cases where $\gamma_i = 0
\;\forall i$ or $\Omega = 0$. Furthermore, there exist only isolated
points in the parameter space such that $\Delta_{\hatLi_R}$ vanishes,
namely those with only one $\gamma_i \neq 0,$ and the corresponding
$\Omega_i=\Omega$ (recall \eqref{eq-ex-real-spherical-coords}). For
all the other choices of the parameters, $\Hi_{\text{DFS}}$ is
attractive by Theorem \ref{thm-attracting-subspace-vec}. Notice that
the Hamiltonian off-diagonal elements are strictly necessary for this
DFS to be attractive, whereas the detuning parameter does not play a
role in determining stability.  As we anticipated in the previous
section, however, the latter may significantly influence the
convergence speed to the DFS for a relevant set in the
parameters-space.
 
\begin{figure}
\centering
\includegraphics[width=7cm]{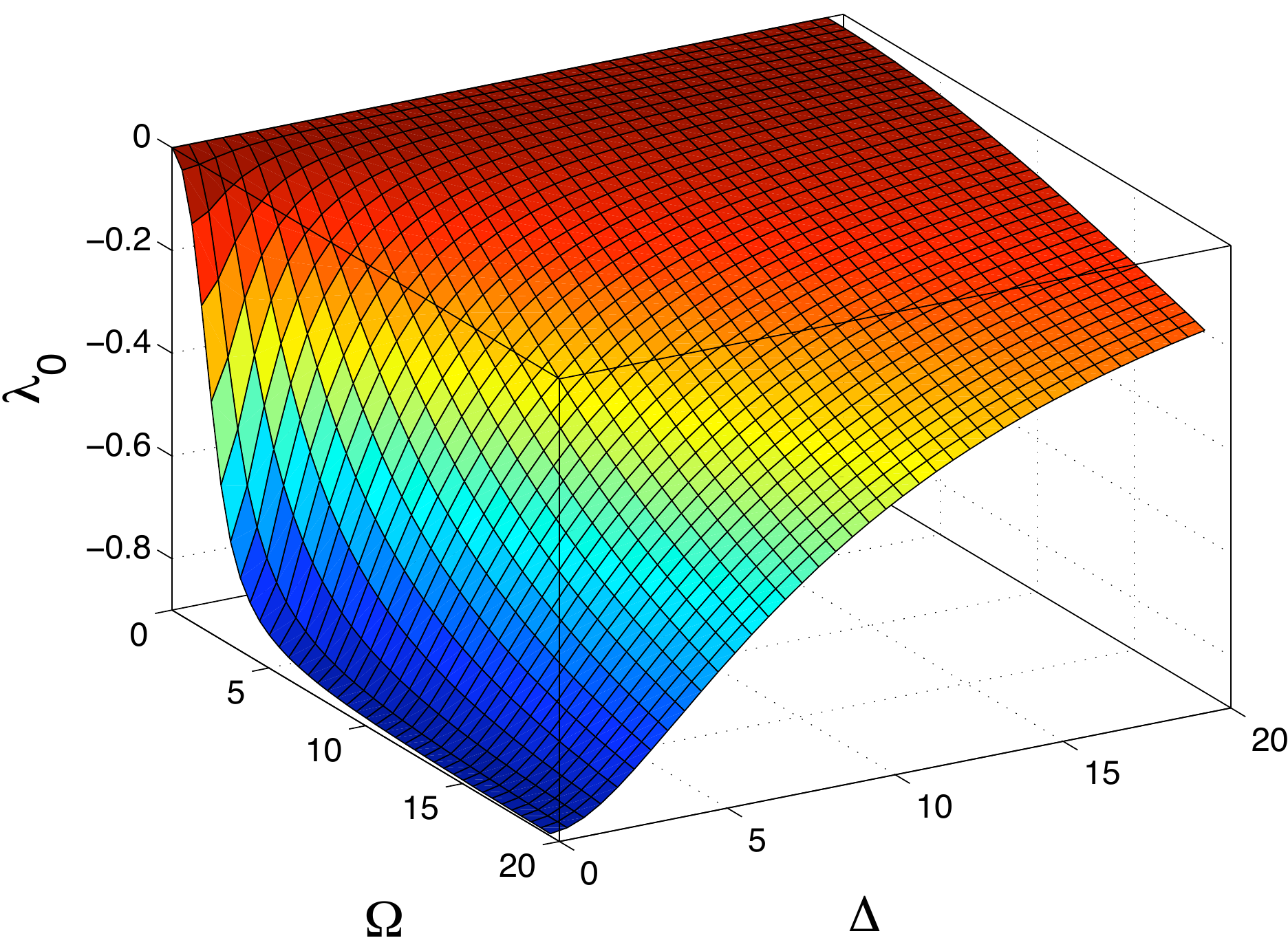}
\caption{Asymptotic convergence speed to the target DFS as a function
    of the parameters $\Delta$ and $\Omega$. We fixed $\gamma_i = 0.9$
    and $\theta=\pi/4$ and $\phi=3/4\pi$. The value of $\lambda_0$ is
    computed by means of Eq. \eqref{speed}.}
\label{fig-ex-real-z0-by_delta-gamma}
\end{figure}

In Figure \ref{fig-ex-real-z0-by_delta-gamma} we graph the value of
$\lambda_0$ given in Eq. \eqref{speed} as a function of $\Delta$ and
$\Omega$, for fixed representative values of $\gamma_i$, $\theta$, and
$\phi$.  As in Example 2, small coupling $\Omega$ as well as high
detuning $\Delta$ slow down the convergence, independently of
$\gamma_i$.  That a strong coupling yields faster convergence directly
reflects the fact that this is fundamental to break the invariance of
the subspace $\span(\ket{r_2})$.  In order to elucidate the effect of
the detuning, consider again the spectrum of $H_R$, which is given by
Eqs. \eqref{spectrumR}- \eqref{eigvecR}.  In the limit of $\Omega
\rightarrow 0$, there exists an eigenvalue $\lambda \rightarrow 0$,
and the same holds for $\Delta \rightarrow \infty$. Furthermore, the
corresponding eigenvector tends to $\ket{r_2}$ in each of these two
limits.  Thus, increasing the detuning can mimic a decrease in the
coupling strength, and vice-versa.

Notice that, unlike in Example 1, there is in this case a non-trivial
dissipative effect linking $\ket{e}$ to $\ket{r_2}$, represented by
the non-zero $R$-blocks of the $L_i$'s, however our design principles
still apply. In fact, the Hamiltonian's off-diagonal terms are still
necessary for $\Hi_{\text{DFS}}$ to be GAS.

\subsection{Dissipative entanglement generation}

The system analyzed in Example 1 is a special instance of a recently
proposed scheme \cite{schirmer-3} for generating (nearly) maximal
entanglement between two identical non-interacting atoms by exploiting
the interplay between collective decay and Hamiltonian tuning.  Assume
that the two atoms are trapped in a strongly damped cavity and the
detuning of the atomic transition frequencies $\omega_i$, $i=1,2$,
from the cavity field frequency $\omega$ can be arranged to be
symmetric, that is, $\omega_1-\omega \equiv \Delta =
-(\omega_2-\omega).$ Under appropriate assumptions \cite{schirmer-3},
the atomic dynamics is then governed by a QDS of the form
\eqref{xsme}, where Eqs. \eqref{entH}-\eqref{entL} are generalized as
follows:
$$L= \sqrt{\gamma} \,(\sigma_+ \otimes I+I\otimes \sigma_+),$$
$$H=\smallfrac{\Delta}{2}(\sigma_z\otimes I - I\otimes \sigma_z)
+\alpha(\sigma_x\otimes I + I\otimes \sigma_x),$$ 
\noindent 
and where, without loss of generality, the parameters $\gamma, \Delta,
\alpha$ may be taken to be non-negative. The QDS still admits an
invariant pure state $\rho_d=|\psi_0\rangle\langle\psi_0|$, which now
depends on the Hamiltonian parameters $\Delta,\,\alpha$:
\[ \ket{\psi_0}= \frac{1}{\Omega}
\Big(\Delta |00\rangle -\alpha (|01\rangle -|10\rangle)\Big), \;\;\;
\Omega=\sqrt {{\Delta}^{2}+2\,{\alpha}^{2}}.\]
\noindent
The DID construction works as in Example 1 (where
$\gamma=\Delta=\alpha=1$), except for the fact that while
$\ket{\psi_1},\ket{\psi_2}$ are defined in the same way as in
\eqref{entB}, the explicit form of fourth basis state \eqref{entB3} is
modified as follows:
\[\ket{\psi_3}= -\frac {1}{\sqrt{2}\Omega}\Big(-2\alpha |00\rangle 
- \Delta (|10\rangle -|01\rangle)\Big). \]
\noindent
Therefore, in matrix representation with respect to the DID basis
$\{\ket{\psi_0},\ket{\psi_2},\ket{\psi_1},\ket{\psi_3}\}$, we get:
\[L= \left[ \begin {array}{c|ccc} 0&{ \sqrt {2}\frac{\Delta}{\Omega}}&0&0\\ 
\hline0&0&\sqrt {2\gamma}&0\\0&0&0&0\\0&-2\,{\frac
{\alpha}{\Omega}}&0&0\end {array} \right],
\]
\[ H= \sqrt{2}\left[ \begin {array}{c|ccc} 0&0&0&0\\ \hline
0&0&\alpha&-\frac {\Omega}{\sqrt {2}}\\0&\alpha&0&0\\0&-\frac
{\Omega}{\sqrt {2}}&0&0\end {array} \right].
\]
The entangled state $\ket{\psi_0}$ is thus GAS. Given the structure of
the above matrices, the following conclusions can additionally be
drawn.  First, the bottleneck to the convergence speed is determined
by the element $L_{12}$, more precisely by the square of the ratio
$\sqrt{2}{\Delta}/{\Omega},$ see \eqref{speed1}. Assuming that $\Delta
\ll \alpha$, the latter is (approximately) linear with the detuning.
This has two implications: on the one hand, the convergence speed
decreases (quadratically) when $\Delta \rightarrow 0.$ On the other
hand, a non-zero detuning is necessary for GAS to be ensured in the
first place: for $\Delta=0,$ the maximally entangled pure state
$\rho_s=\ket{\psi_s}\bra{\psi_s}$, with
$\ket{\psi_s}=\smallfrac{1}{\sqrt{2}}(\ket{01}-\ket{10})$, cannot be
perfectly stabilized.
Likewise, although the parameter $\alpha$ plays no key role in
determining GAS, a non-zero $\alpha$ is nevertheless fundamental in
order for the asymptotically stable state $|\psi_0\rangle$ to be
entangled.

\subsection{State preparation in coupled electron-nuclear systems}

We consider a bipartite quantum system composed by nuclear and
electronic degrees of freedom, which is motivated by the well-known
Nitrogen-Vacancy (NV) defect center in
diamond~\cite{Jacques09,enhancedNV,Jiang09,Neumann10b}.  While in
reality both the electronic and nuclear spins (for $^{14}N$ isotopes)
are spin-1 (three-dimensional) systems, we begin by discussing a
reduced description which is common when the control field can address
only selected transitions between two of the three physical
levels.  The full three-level system will then be discussed at the end
of the section.

\subsubsection{Reduced model}

Let both the nuclear and the electronic degrees of freedom be described as spin 1/2
particles.  In addition, assume that the electronic state can transition from
its energy ground state to an excited state through optical pumping in
a spin-preserving fashion.  The decay from the excited state, on the
other hand, can be either spin preserving or temporarily populate a
{\em metastable state} from which the electronic spin decays only to
the spin state of lower energy~\cite{Manson06}.  We describe the
overall optically-pumped dynamics of the NV system by constructing a
QDS generator. A basis for the reduced system's state space is given
by the eight states
$$\ket{E_{el},s_{el}}\otimes\ket{s_N} \equiv
|E_{el},s_{el},s_N\rangle,$$
\noindent 
where the first tensor factor describes the electronic degrees of
freedom, that are specified by the energy levels $E_{el}=g,e,$ and the
electron spin $s_{el}=0,1$ (corresponding to the spin pointing up or
down, respectively), while the second factor refers to the nuclear
spin and can take the values $s_N=0,1$.  To these states one has to
add the two states belonging to the metastable energy level, denoted
by $\ket{ms}\otimes\ket{s_N},$ with $s_N$ as before.  Notice that a
``passage'' through the metastable state erases the information on the
electron spin, while it does preserves the nucleus spin state.

The Hamiltonian for the coupled system is of the form
$H_{tot}=H_g+H_e,$ where the excited-state Hamiltonian $H_e$ and the
ground-state Hamiltonian $H_g$ share the following structure: \beqa
\label{hamstruct}
H_{g,e}&=& D_{g,e} S_z^2\otimes \identity_N +Q\ \identity_{el}\otimes
S_z^2 \nonumber \\ && + B\,( g_{el} S_z \otimes \identity_N+g_n
\identity_{el}\otimes S_z )\\ &&+\frac{A_{g,e}}2 ( S_x\otimes S_x +S_y
\otimes S_y+2S_z \otimes S_z ).\nonumber \eeqa 
\noindent 
Here, $S_{x,y}=\sigma_{x,y},$ are the standard $2\times2$ Pauli
matrices on the relevant subspace, while
$S_z=\tfrac{1}{2}(\identity-\sigma_z)$ is a pseudo-spin\footnote{This
different definition follows from the implemented reduction from a
three- to a two- level system: specifically, we consider only
$\ket{0,-1}$ and neglect $\ket{1}$, and further map the states $0\to0$
and $-1\to1$.}  and $D_{g,e},A_{g,e}, Q$ are fixed parameters. In
particular, $A_{g,e}$ will play a key role in our analysis,
determining the strength of the Hamiltonian ({\em hyperfine})
interaction between the electronic and the nuclear degrees of
freedom. On the other hand, $B$ represents the intensity of the
applied static magnetic field along the $z$-axis, and can be thought
as the {control parameter}.

In order to describe the non-Hamiltonian part of the evolution we
employ a phenomenological model, using Lindblad terms with jump-type
operators and associated pumping and decay rates.  The relevant
transitions are represented by the operators (including the decay
rates) below: since they leave the nuclear degrees of freedom
unaltered, they act as the identity operator on that tensor factor.
Specifically:
\beq
\label{LopNV}
\begin{array}{lcl}
 L_{1}&=& \sqrt{\gamma_d}\;\ket{g,0}\bra{e,0}\otimes \identity_N, \\
 L_{2}&=& \sqrt{\gamma_d}\;\ket{g,1}\bra{e,1}\otimes \identity_N,\\
  L_{3}&=& \sqrt{\gamma_m} \; \ket{ms}\bra{e,1}\otimes \identity_N,\\ 
 L_{4}&=& \sqrt{\gamma_0} \; \ket{g,0}\bra{ms}\otimes \identity_N,\\
L_{5}&=&\sqrt{\gamma_p} \; \ket{e,0}\bra{g,0}\otimes \identity_N,\\
 L_{6}&=&\sqrt{\gamma_p} \; \ket{e,1}\bra{g,1}\otimes \identity_N.
 \end{array}
\eeq 
\noindent 
The first four operators describe the decays, with associated rates $\gamma_{d},\,\gamma_{m}\, \gamma_{0},$ whereas the last two
operators correspond to the optical-pumping action on the electron,
with a rate $\gamma_{p}.$ It is easy to check by inspection that the
subspace
$$\Hi_S: =\textrm{span}\{\ket{e,0,0},\ket{g,0,0}\}$$ is invariant for
the dissipative part of the dynamics: we next establish it is also
GAS, and analyze the dynamical structure induced by the
DID.

\paragraph{Convergence analysis} Following the procedure presented
in Sec. \ref{sectionDID}, we can prove that $\Hi_S$ is
attractive. This is of key interest in the study of NV-centers as a
physical platform for solid-state quantum information processing. In
fact, it corresponds to the ability to perfectly {\em polarize} the
joint spin state of the electron-nucleus system\footnote{Note that  non-spin conserving decay and pumping processes~\cite{Manson06}, which we neglected here, limit in practice the achievable polarization.}. The proposed DID
algorithm runs to completion (in seven iterations), with the following
basin decomposition as output:
$$\Hi=\Hi_S\oplus\Hi^{(1)}_T\oplus\ldots\oplus\Hi^{(7)}_T,$$
where
\beqan 
\Hi^{(1)}_T&=&\textrm{span}\{\ket{ms,0}\},\\
\Hi^{(2)}_T&=&\textrm{span} \{\ket{e,1,0}\},\\
\Hi^{(3)}_T&=&\textrm{span} \{\ket{g,1,0}\},\\
\Hi^{(4)}_T&=&\textrm{span} \{\ket{e,0,1},\ket{g,0,1}\},\\
\Hi^{(5)}_T&=&\textrm{span} \{\ket{ms,1}\},\\
\Hi^{(6)}_T&=&\textrm{span} \{\ket{e,1,1}\},\\
\Hi^{(7)}_T&=&\textrm{span} \{\ket{g,1,1}\}.
\eeqan
\noindent 
We do not report the block form of every operator, both for the sake
of brevity and because it would not be very informative: we report
instead the {\em Dynamical Connection Matrix} (DCM) associated to the
above DID. The latter is simply defined as $C:=H_{tot}+\sum_k L_k,$
with all the matrices represented in the energy eigenbasis, ordered
consistently with the DID. While in general the DCM would not provide
sufficient information to determine the invariance of the various
subspaces due to the fine-tuning conditions derived in
\eqref{invarianceconditions}, having only $L_k$ of jump-type (or
creation/annihilation) greatly simplifies the analysis. In fact, it is
easy to see that a non-zero entry $C_{ij}$ due to some $L_k$ means
that the $j$-th state of the basis is attracted towards the $i$-th
one. Thus, the DCM gives a compact representation of the dynamical
connections between the basins, pointing to the available options for
Hamiltonian tuning. Explicitly:
\[\label{DCMNV}
C=\hspace{-2mm}\left[
\begin{array}{cc|c|c|c|cc|c|c|c}
\cellcolor[gray]{0.85} 0 &\cellcolor[gray]{0.85} \gamma_{p}^{\frac{1}{2}} &  &  &  &  &  &  & & \\ 
 \cellcolor[gray]{0.85}\gamma_d^{\frac{1}{2}} &\cellcolor[gray]{0.85} 0 & \gamma_0^{\frac{1}{2}} &  &  &  &  &  & &\\\hline 
  &  &\cellcolor[gray]{0.85} 0& \gamma_m^{\frac{1}{2}} &  &  &  &  & & \\\hline 
  &  &  &\cellcolor[gray]{0.85} h_e & \gamma_p^{\frac{1}{2}} & A_e & 0 &  & & \\    \hline 
  &  &  & \gamma_d^{\frac{1}{2}} &\cellcolor[gray]{0.85} h_g & 0 & A_g &  & & \\\hline 
  &  &  & A_e & 0 &\cellcolor[gray]{0.85} h_n & \cellcolor[gray]{0.85}\gamma_p^{\frac{1}{2}} &  & &\\
  &  &  & 0 & A_g &\cellcolor[gray]{0.85} \gamma_d^{\frac{1}{2}} &\cellcolor[gray]{0.85} h_n & \gamma_0^{\frac{1}{2}} & &\\\hline 
  &  &  &  &  &  &  & \cellcolor[gray]{0.85}h_n & \gamma_m^{\frac{1}{2}} &\\  \hline 
  &  &  &  &  &  &  &  &\cellcolor[gray]{0.85} h'_e & \gamma_p^{\frac{1}{2}} \\  \hline 
  &  &  &  &  &  &  &  & \gamma_d^{\frac{1}{2}} &\cellcolor[gray]{0.85} h'_g\\  
  \end{array}
\right]
\]
with $h_{e,g}:=D_{e,g}- {g_{el}}B$, $h'_{e,g}:=D_{e,g}-
({g_{el}}+{g_n})B+Q+A_{e,g}$, and $h_n:= Q-g_nB$.
By definition, the block division (highlighted by the solid lines) is
consistent with the DID, and all the empty blocks are zero. In some
sense, the DCM can be seen as a ``graph-based'' visualization, similar
to those commonly use to determine the controllability properties of
closed quantum systems
\cite{altafini-controllability,turinici-controllability}.

Since in our problem $g_n \ll g_{el}$, the diagonal entries in the
Hamiltonian that are most influenced by the control parameter $B$ are
$h_{g,e}$ and $h'_{g,e}$.  All the other entries of the DCM are (for
typical values of the parameters) either independent or only very
weakly dependent on $B$.

It is immediate to see that the $\gamma_p,\gamma_m,\gamma_0$ blocks
establish dynamical connections between all the neighboring basins,
with the exception of $\Hi_T^{(4)}$ which is connected by the
($B$-independent) Hamiltonian elements $A_e,\,A_g$ to $\Hi_T^{(2)}$,
$\Hi_T^{(3)}$.  The DCM also confirms the fact that $\Hi_S$ is
invariant, since its first column, except the top block, is zero.  In
the terminology of Sec. \ref{basins}, $\Hi_T^{(1)}$ is the only
transition basin, $\Hi_T^{(4)}$ is the only circulant basin, and all
the other basins are mixing basins.  It is worth remarking that {\em
any choice of the control parameter $B$ ensures GAS}. By inspection of
the DCM, one finds that the bottleneck in the noise-induced
connections between the basins is determined by the
$\gamma_0,\gamma_p$ parameters. Since the latter are not affected by
the control parameter, the minimum of those rates will determine the
fundamental limit to the speed of convergence to $\Hi_S$ in our
setting. 

\paragraph{Optimizing the convergence speed}
The only transitions which are significantly influenced by $B$ are the
ones connecting $\Hi_T^{(4)}$ to $\Hi_T^{(2)}$ and $\Hi_T^{(3)}$. By
appropriately choosing $B$ one can reduce the norm of $h_e$ or $h_g$
to zero, mimicking ``resonance'' condition of Example 2.  Assume that,
as in the physical system, $A_e>A_g$. Considering that $\Hi_T^{(2)},$
associated to $h_e,$ is coupled to $\Hi_T^{(4)}$ with the largest
off-diagonal Hamiltonian term ($A_e$) and it is closer to $\Hi_S$ in
the DID, we expect that the best performance will be obtained by
imposing $h_e\equiv 0$, that is, by setting $B\equiv D_e/g_{el}$.

\begin{figure}[t]
\centering
\includegraphics[width=8.5cm,height=8cm]{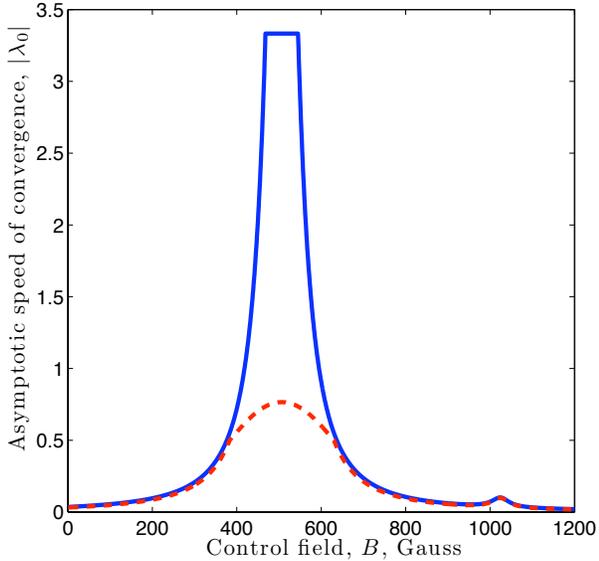}
  \caption{Asymptotic convergence speed to $\Hi_S$ as a function of
    the control parameter $B$ for an NV-center. The blue (solid) curve
    is relative to the system with the metastable state, while the red
    (dashed) one is relative to a simplified system where the
    transition through the metastable state has been incorporated in a
    single decay operator $\tilde L$ with rate $\gamma_0$ (see text).
    Typical values for NV-centers are: $D_e=1420$ MHz, $D_g=2870$MHz,
    $Q=4.945$MHz, $A_e=40$ MHz, $A_g=2.2$MHz and $g_{el}=2.8$MHz/G,
    $g_n=3.08\times10^{-4}$MHz/G.  We used decay rates $\gamma_{d}=77
    \textrm{MHz},\, \textrm{MHz},\,\gamma_{m}=33 \textrm{MHz},\,
    \gamma_{0}=3.3 \textrm{MHz}$, and optical-pumping rate
    $\gamma_{p}=70 \textrm{MHz}.$ With these values, $h_g'\approx
    h_g=2870-2.8B$ MHz, $h_e'\approx h_e=1420-2.8B$ MHz and
    $h_n\approx4.945$ MHz.}
\label{speedinB}
\end{figure}

The above qualitative analysis is confirmed by numerically computing
the exact asymptotic convergence speed, Eq. \eqref{speed}.  The
behavior as a function of $B$ is depicted in Figure \ref{speedinB}.
It is immediate to notice that the maximum speed is indeed limited by
the lowest decay rate, that is, the lifetime $\gamma_0$ of the singlet
state with our choice of parameters. The maximum is attained for
near-resonance control values, although exact resonance,
$B=D_e/g_{el}$, is actually not required. The second (lower) maximum
correspond to the weaker resonance that is attained by choosing $B$ so
that $h_g=0$. Physically, tuning the control parameter such that
$h_e=0$ corresponds, in our reduced model, the excited-state ``level
anti-crossing'' (LAC) condition that has been experimentally
demonstrated in \cite{Jacques09}.

In Figure \ref{speedinB}, we also plot (dashed line) the speed of
convergence of a simplified system where the transition through the
metastable state and its decay to the ground state have been
incorporated in a single decay operator $\tilde L=L_4L_3,$ with a rate
equal to $\gamma_0$.  This may seem convenient, since once the decay
to the metastable level has occurred, the only possible evolution is a
further decay into $\ket{g,1,1}.$ However, by doing this the overall
convergence speed appears to be substantially diminished, although
still in qualitative agreement with the predicted behavior (presence
of the two maxima, and speed lower than the minimum decay rate). The
reason lies in the fact that in this reduced model, the
$\Hi_T^{(1)},\Hi_T^{(5)}$ transition basins become (part of) mixing
basins, thus the non-polarizing decay and the Hamiltonian can directly
influence (slowing down) the decay dynamics associated to $\tilde L$.

\subsubsection{Extended model and practical stabilization}

A physically more accurate description of the NV-center system
requires representing both the electron and nucleus subsystems as
three-level, spin-$1$ systems.  In this case, a basis for full 
state space is given by the 21 states
$$\ket{E_{el},s_{el}}\otimes\ket{s_N},\;\ket{ms}\otimes\ket{s_N},$$ 
\noindent 
where now the electron spin can take values $s_{el}=1,0,-1$ and,
similarly, the nucleus spin is labeled by $s_N=1,0,-1$. The
Hamiltonian is of the form: \beqa
\label{hamsspin1} H_{g,e}&=& D_{g,e} S_z^2\otimes
\identity_N +Q\ \identity_{el}\otimes S_z^2 \nonumber \\ && + B\,(
g_{el} S_z \otimes \identity_N+g_n \identity_{el}\otimes S_z )\\
&&+{A_{g,e}} ( S_x\otimes S_x +S_y \otimes S_y+S_z \otimes S_z
),\nonumber \eeqa 
\noindent 
where now $S_{x,y,z}$ are the angular momentum operators for
spin-1. The non-Hamiltonian part of the evolution is given by the
operators in Eq. \eqref{LopNV} (where now $0,1$ correspond to the
spin-1 eigenstates), to which one needs to add the dissipation
channels associated to Lindblad operators: \beqan L_{7}&=&
\sqrt{\gamma_d}\;\ket{g,-1}\bra{e,-1}\otimes \identity_N,\\ L_{8}&=&
\sqrt{\gamma_m} \; \ket{ms}\bra{e,-1}\otimes \identity_N,\\
L_{9}&=&\sqrt{\gamma_p} \; \ket{e,-1}\bra{g,-1}\otimes \identity_N.
\eeqan 
\noindent 
Similar to the spin 1/2 case, the dissipative dynamics alone would
render GAS the subspace associated to electronic spin $s_{el}=0$,
that is $\Hi_{el,0}={\rm span}\{\ket{e,0,s_n},\ket{g,0,s_n}\}$.  Thus,
one may hope that $\Hi_S=\rm{span}\{\ket{e,0,0},\ket{g,0,0}\}$ could
still be GAS under the full dynamics. We avoid reporting the whole DID
and the DCM structure, since that would be cumbersome and unnecessary
to our scope: the main conclusion is that in this case {\em nuclear
spin polarization cannot be perfectly attained.} While the hyperfine
interaction components of the Hamiltonian still effectively connect
the subspaces with nuclear spin $0,1$, they also have a detrimental
effect: $\Hi_S$ is no longer invariant. In fact, represented in the
DID basis, the Hamiltonian has the following form:
\[
H_{tot}= \left[ \begin {array}{cc|ccccc} 
 0&0&0&\cdots&A_e&0&\cdots\\
 0&0&0&\cdots&0&A_g&\cdots\\
 \hline
 0&0&0&\cdots&0&0&\cdots\\
\vdots&\vdots&\vdots&\ddots&\vdots&\vdots&\ddots\\
{\it A_e}&0& 0&\cdots &0&0&\cdots\\
0&{\it A_g}& 0&\cdots &0&0&\cdots\\
\vdots&\vdots&\vdots &\ddots &\vdots&\vdots&\ddots\\
 \end {array} \right]. 
\]

\noindent The presence of $A_g,A_e$ in the $H_P^\dag$ block suffices to {\em
destabilize} $\Hi_S,$ by causing the invariance conditions in
\eqref{invarianceconditions} to be violated.  However, these terms are
relatively small compared to the dominant ones, allowing for a
practical stabilization attempt. Following the approach outlined in
Sec. \ref{practical}, we neglect the $H_P$ term and proceed with the
analysis and the convergence-speed tuning. Again, the optimal speed
condition is attained for $B$ in a nearly-resonant LAC condition. By
means of numerical computation, one can then show that the system
still admits a {\em unique}, and hence attractive, steady state (which
in this case is mixed) and that the latter is close to the desired
subspace. In fact, with the same parameters we employed in the
spin-1/2 example, one can ensure asymptotic preparation of a state
with polarized $s_N=1$ spin with a fidelity of about 97\%.


\section{Conclusions}

We have developed a framework for analyzing global asymptotic
stabilization of a target pure state or subspace (including practical
stabilization when exact stabilization cannot be attained) for
finite-dimensional Markovian semigroup driven by time-independent
Hamiltonian controls.  A key tool for verifying stability properties
under a given semigroup generator is provided by a canonical
state-space decomposition into orthogonal subspaces, uniquely
determined by the effective Hamiltonian and Lindblad operators (the
DID), for which we have provided a constructive algorithm and an
enhanced version that can accommodate control constraints.  In the
second part of the work, we have tackled the important practical
problem of characterizing the speed of convergence to the target
stable manifold and the extent to which it can be manipulated by
Hamiltonian control parameters. A quantitative system-theoretic lower
bound on the attainable speed has been complemented by a
hydraulic-inspired connected-basin approach which builds directly on
the DID and, while qualitative, offers more transparent insight on the
dynamical effect of different control knobs.  In particular, such an
approach makes it clear that even control parameters that have no
effect on invariance and/or attractivity properties may significantly
impact the overall convergence speed.

While our results are applicable to a wide class of controlled
Markovian quantum systems, a number of open problems and extensions
remain for future investigation.
From a theory standpoint, it could be interesting to study the speed of
convergence to the system's equilibria or limit sets without {\em a
priori} knowing their structure.  To this end, one could envision to
first determine the minimal collector basin of the dynamics, and then
analyze stability with that as a target
(note that the fact the minimal collector basin contains the target
subspace does not ensure invariance of the latter).
Likewise, for practical applications, an important question is whether
similar analysis tools and design principles may be developed for more
general classes of controls than addressed here.  In this context, the
case where a set of tunable Lindblad operators may be applied
open-loop, alone and/or in conjunction with time-independent
Hamiltonian control,
may be especially interesting, and potentially relevant to settings
that incorporate engineered dissipation and {\em dissipative gadgets},
such as nuclear magnetic resonance \cite{violaHavel} or trapped-ion
and optical-lattice quantum simulators \cite{zoller,cirac}.

\bibliography{bibliography}

\bibliographystyle{IEEEtran}

\end{document}